%% file: Jozwiak_TI_total.tex
\documentclass[prb,twocolumn,superscriptaddress,floatfix,letterpaper]{revtex4-1}

\usepackage{graphicx,amssymb,bm,amsfonts,natbib,amsmath,graphics,color}

\usepackage[bookmarks=false,pdfstartview=FitH,colorlinks=true,citecolor=blue,linkcolor=blue]{hyperref}

\usepackage{chapterbib}

\usepackage{setspace}

\begin{document}
\include{Jozwiakmain}

\renewcommand{\theequation}{s\arabic{equation}}
\renewcommand{\thefigure}{S\arabic{figure}}
\renewcommand{\thesection}{SI \Roman{section}}
\renewcommand{\thepage}{SI \arabic{page}}

\setcounter{page}{1}
\setcounter{figure}{0}
\setcounter{equation}{0}

\begin{widetext}
\begin{large}

\include{Jozwiaksi}
\end{large}
\end{widetext}
\end{document}

%% file: Jozwiakmain.tex
%
%

\title {Photoelectron spin-flipping and texture manipulation in a topological insulator}

\author{Chris Jozwiak} \affiliation{Advanced Light Source, Lawrence Berkeley National Laboratory, Berkeley, California 94720, USA}
\author{Cheol-Hwan Park} \affiliation{Department of Physics, University of California, Berkeley, California 94720, USA} \affiliation{Department of Physics and Astronomy and Center for Theoretical Physics, Seoul National University, Seoul 151-747, Korea}
\author{Kenneth Gotlieb} \affiliation{Graduate Group in Applied Science \& Technology, University of California, Berkeley, California 94720, USA}
\author{Choongyu Hwang} \affiliation{Materials Sciences Division, Lawrence Berkeley National Laboratory, Berkeley, California 94720, USA}
\author{Dung-Hai Lee} \affiliation{Department of Physics, University of California, Berkeley, California 94720, USA} \affiliation{Materials Sciences Division, Lawrence Berkeley National Laboratory, Berkeley, California 94720, USA}
\author{Steven G. Louie} \affiliation{Department of Physics, University of California, Berkeley, California 94720, USA} \affiliation{Materials Sciences Division, Lawrence Berkeley National Laboratory, Berkeley, California 94720, USA}
\author{Jonathan D. Denlinger} \affiliation{Advanced Light Source, Lawrence Berkeley National Laboratory, Berkeley, California 94720, USA}
\author{Costel R. Rotundu} \affiliation{Materials Sciences Division, Lawrence Berkeley National Laboratory, Berkeley, California 94720, USA}
\author{Robert J. Birgeneau} \affiliation{Department of Physics, University of California, Berkeley, California 94720, USA} \affiliation{Materials Sciences Division, Lawrence Berkeley National Laboratory, Berkeley, California 94720, USA} \affiliation{Department of Materials Science and Engineering, University of California, Berkeley, California 94720, USA}
\author{Zahid Hussain} \affiliation{Advanced Light Source, Lawrence Berkeley National Laboratory, Berkeley, California 94720, USA}
\author{Alessandra Lanzara} \affiliation{Department of Physics, University of California, Berkeley, California 94720, USA} \affiliation{Materials Sciences Division, Lawrence Berkeley National Laboratory, Berkeley, California 94720, USA}

\maketitle

\textbf{
Recently discovered materials called three-dimensional topological insulators\cite{Fu2007,Moore2007,Roy2009,Qi2010,Moore2010} constitute examples of symmetry protected topological states in the absence of applied magnetic fields and cryogenic temperatures.
A hallmark characteristic of these non-magnetic bulk insulators is the protected metallic electronic states confined to the material's surfaces.
Electrons in these surface states are spin polarized with their spins governed by their direction of travel (linear momentum), resulting in a helical spin texture in momentum space.\cite{Hsieh2009a}
Spin- and angle-resolved photoemission spectroscopy (spin-ARPES) has been the only tool capable of directly observing this central feature with simultaneous energy, momentum, and spin sensitivity.\cite{Hsieh2009,Hsieh2009a,Nishide2010,Pan2011a,Souma2011a,Xu2011a,Jozwiak2011}
By using an innovative photoelectron spectrometer\cite{Jozwiak2010} with a high-flux laser-based light source, we discovered another surprising property of these surface electrons which behave like Dirac fermions.
We found that the spin polarization of the resulting photoelectrons can be fully manipulated in all three dimensions through selection of the light polarization. 
These surprising effects are due to the spin-dependent interaction of the helical Dirac fermions with light, which originates from the strong spin-orbit coupling in the material.
Our results illustrate unusual scenarios in which the spin polarization of photoelectrons is completely different from the spin state of electrons in the originating initial states.
The results also provide the basis for a novel source of highly spin-polarized electrons with tunable polarization in three dimensions.}

The topological electronic bandstructure of a bulk topological insulator ensures the presence of gapless surface electronic states with Dirac-like dispersions similar to graphene.
Unlike graphene, the topological surface states are spin polarized, with their spins locked perpendicular to their momentum, forming helical spin-momentum textures\cite{Hsieh2009a} (see Fig. 1(a)).
The presence of such `helical Dirac fermions' forms an operational definition of a 3D topological insulator, and much of the excitement surrounding topological insulators involves the predicted exotic phenomena and potential applications of these metallic states.\cite{Qi2010,Moore2010}
These include novel magnetoelectric effects,\cite{Qi2008} exotic quasiparticles (in a proximity induced superconducting state) called Majorana fermions which are their own antiparticles,\cite{Fu2008} and applications ranging from spintronics to quantum computing.\cite{Nayak2008}
Establishing methods that are sensitive to these states and their predicted behaviors have therefore generated much interest.\cite{Hsieh2009a,Roushan2009,Peng2010,Hsieh2011,McIver2012}

\begin{figure*} \includegraphics[width=18cm]{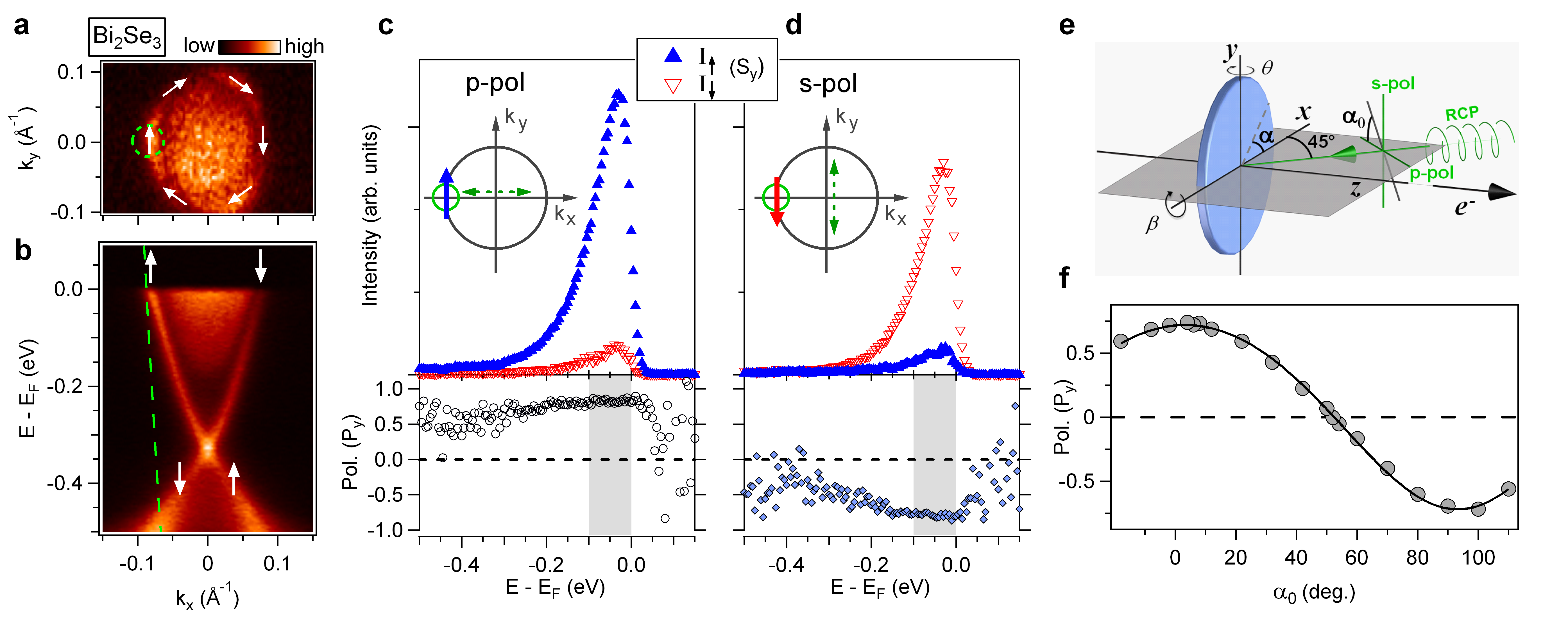}
\caption{\label{fig:fig1} \textbf{The dependence of photoelectron spin on linear photon polarization observed in a topological insulator.}  
(a) ARPES intensity map at $E_F$ of the (111) surface of Bi$_2$Se$_3$, with the $\Gamma$M direction aligned along $k_x$.  The white arrows show the expected spin polarization around the surface state Fermi surface.
(b) ARPES intensity map as a function of binding energy and momentum.
(c) Spin-resolved photemission intensity as a function of binding energy, at fixed emission angle ($\sim$ fixed $\mathbf{k}$), corresponding to the dashed line-cut in (b), and the momentum location marked by the dashed circle in (a).  The corresponding $y$ component of the photoelectron polarization, $P_y$, is shown in the bottom panel.  The Fermi surface diagram inset highlights the k-space location, ($k_x$,$k_y$)=($-k_\textrm{F}$,$0$) (green circle), along with the spin-polarization direction indicated by the data.  The data are acquired with $p$-polarized photons, with the photon polarization vector, projected into the sample surface plane, shown as a dashed green arrow in the inset.
(d) Same as (c), but with $s$-polarized photons.
(e) Diagram of the experimental geometry.   Linear polarization of photons can be continuously rotated as shown.  Dashed gray line represents projection of incident light linear polarization on the sample surface.
(f) Photoelectron spin polarization at ($k_x$,$k_y$)=($-k_\textrm{F}$,$0$) as a function of rotation of the photon polarization.  Photoelectron polarization is integrated in binding energy corresponding to the gray regions of the bottom panels in (c) and (d).  Black curve is a fit following the presented theory (see Supplemental Information).
}
\end{figure*}

Angle-resolved photoemission spectroscopy (ARPES) directly maps the dispersions and Fermi surfaces of such electronic states in energy-momentum space.
Spin-resolved ARPES also measures the spin polarization of the corresponding photoelectrons.
Following a common assumption that electron spin is conserved in the photoemission process, the technique has been used to identify the presence of the predicted helical spin textures of topological surface states.\cite{Hsieh2009,Hsieh2009a,Nishide2010,Pan2011a,Souma2011a,Xu2011a,Jozwiak2011}
Utilizing a high-efficiency spin-resolved photoelectron spectrometer\cite{Jozwiak2010} and a high intensity laser light source that enabled rapid high-resolution data acquisition, we have found surprising new features of the photoelectron spin texture in the prototypical topological insulator, Bi$_2$Se$_3$.\cite{Zhang2009}
In particular, the results demonstrate strong dependence of the photoelectron spin polarization on the photon polarization, enabling its full manipulation.
This dramatically illustrates that spin-conservation, commonly assumed for photoemission, is invalid in these materials.\cite{CheolHwan}

Figures~1(a,b) show standard ARPES data collected from a Bi$_2$Se$_3$ single crystal. 
The sharp surface states form a cone-like dispersion in panel (b), characterized by the ring-like Fermi surface piece in panel (a).
The sample is n-doped, such that the bottom of the bulk conduction band falls below the Fermi level, forming the chunk of spectral weight in the center of the surface state cone.\cite{Xia2009,Jozwiak2011}

Figures~1(c)~and~(d) show spin-resolved energy distribution curves (EDCs), or plots of photoelectron intensity as a function of binding energy at a particular momentum, corresponding to the line-cut marked in panel (b).
The EDCs are resolved into distinct channels for spin-up and -down photoelectrons.
Here, the spin quantization axis is the $y$-axis.
The corresponding spin polarization, or $P_y$, curves are shown below, and are a measure of the relative difference between the number of spin-up and -down photoelectrons according to $P_y = \frac{I_\uparrow - I_\downarrow}{I_\uparrow + I_\downarrow}$.
The data were acquired with linearly polarized light in two distinct photon polarization geometries, in which  the electric field vector, $\hat{\epsilon}$, was in the $xz$-plane ($p$-polarization) and along the $y$-axis ($s$-polarization), respectively (see Fig.~1(e) and insets in panels c-d).

In the case of $p$-polarized light (panel c), the intensity peak is primarily spin-up, and its $P_y$ is nearly $+1$.
Thus, photoelectrons from the surface state near $E_F$ with momentum ($k_x,k_y$) = (-$k_F$,0) were strongly polarized `up' along the positive $y$-axis as labeled by the blue arrow in the inset. 
This is consistent with previous spin-ARPES measurements \cite{Hsieh2009,Hsieh2009a,Nishide2010,Pan2011a,Souma2011a,Xu2011a,Jozwiak2011} and with the predicted helical spin texture (Fig.~1(a)) where the surface state spins are tangential to the Fermi surface contour with clockwise helicity.
Remarkably, when the light polarization is rotated by ${\pi/2}$ to $s$-polarization (panel d), the intensity peak reversed to primarily spin-down, with its $P_y$ nearly $-1$, such that photoelectrons from the same initial state were polarized `down' as labeled by the red arrow in the inset (panel d).
This is opposite to the expected spin texture for the surface states.
Taken on its own, this result would counter previous spin-ARPES results, likely taken only with $p$-polarized light,\cite{Hsieh2009a,Pan2011a,Jozwiak2011} and seem to point to a spin texture of reversed helicity.

Furthermore, Fig.~1(f) shows the corresponding measured photoelectron polarization component, $P_y$, as a function of continuous rotation of the photon polarization vector between $p$- and $s$-polarizations.
Clearly, the photoelectron spin polarization is dependent on the photon polarization and can be continuously modulated from nearly $P_y=+1$ to $-1$.

\begin{figure*} \includegraphics[width=13cm]{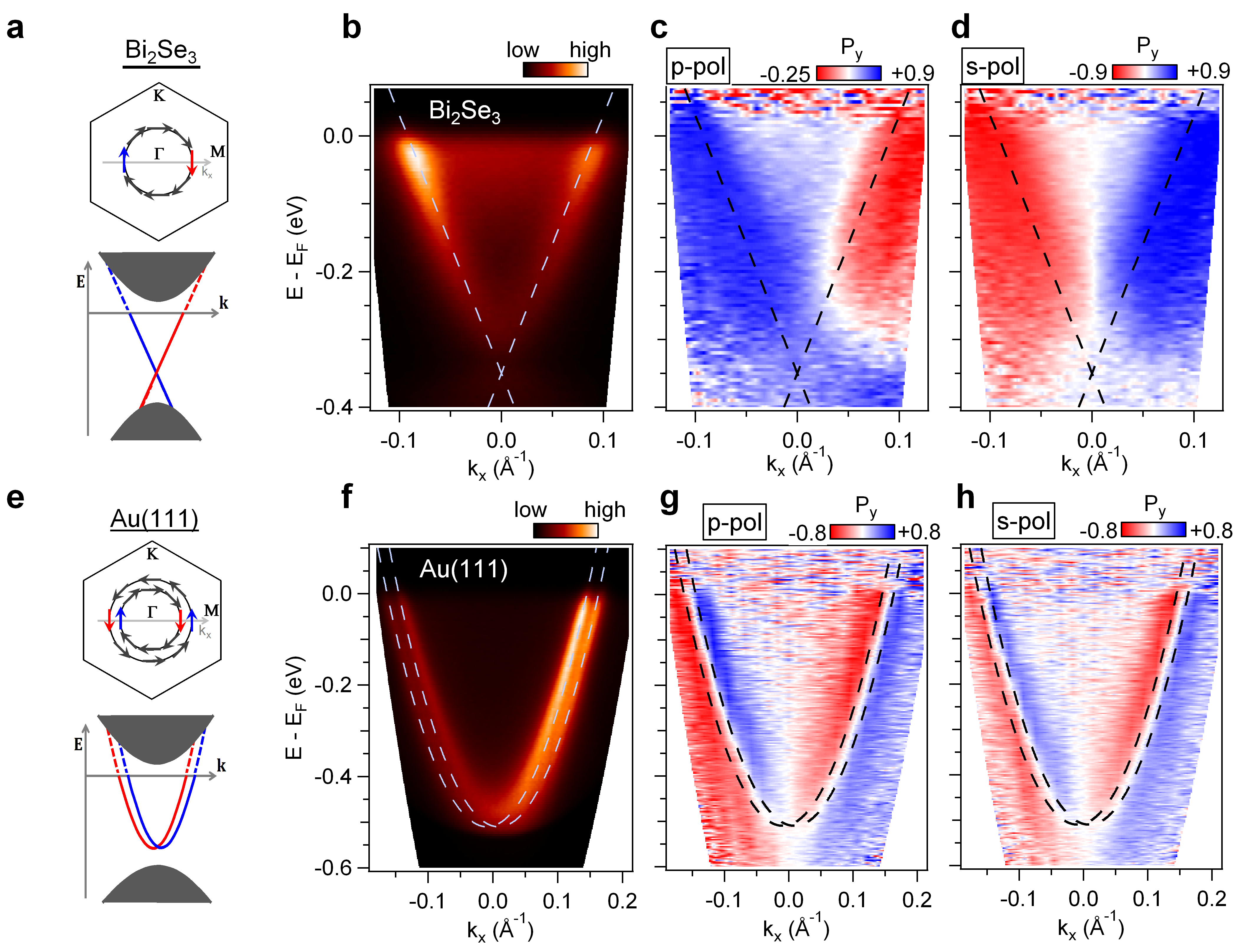}
\caption{\label{fig:cores} \textbf{Photoelectron spin flipping mapped through momentum space.}
(a) Schematic of surface state helical Dirac fermions in Bi$_2$Se$_3$, including Fermi surface (above) and energy dispersion along $k_x$.
(b) Spin-integrated ARPES intensity map of Bi$_2$Se$_3$, taken with laser, s-polarized, h$\nu$ = 5.99 eV.  Dashed lines are linear guides to the eye illustrating Dirac cone dispersion of the surface state.
(c,d) Corresponding spin polarization ($P_y$) maps taken with p- and s-polarized light, respectively.  Dashed guides to the eye are identical to (b).
(e-h) Same as a-d, but for the Au(111) surface state.  Dashed lines in f-h are parabolic guides to the eye following the free-electron-like dispersions.
}
\end{figure*}

The full energy and momentum dependence of this photoelectron spin flipping in Bi$_2$Se$_3$ is shown in Fig.~2 and is compared with the Rashba spin-split Au(111) Shockley surface state.\cite{Hoesch2004}
For reference, schematics of the theoretical spin-polarized surface band dispersion and Fermi surface spin texture of the Bi$_2$Se$_3$ and Au(111) surface states are shown in panels~(a,e), and the measured spin-integrated ARPES maps of the corresponding photoemitted electrons as a function of binding energy and $k_x$ are shown in panels~(b,f), respectively.
Panels~(c,d) (and (g,h)) show the corresponding complete photoelectron $P_y$ maps for $p$- and $s$-polarized light, respectively, for the Bi$_2$Se$_3$ (and Au) surface state. 
In both cases,  when the light is $p$-polarized, the photoelectron spin texture matches the expected surface state spin texture (compare panel (c) with (a) and panel (g) with (e)).
Specifically, for Bi$_2$Se$_3$, the photoelectrons following the branch of the Dirac cone with negative slope are `spin-up' (blue), and those along the branch with positive slope are `spin-down' (red), as expected.
Panel~(c) is shown with an asymmetric color scale to offset an overall shift in $P_y$ due to its particular experimental geometry (see Supplementary Information).
Similarly for the Au(111) surface state in panel (g), the photoelectrons corresponding to the nearly-free electron parabola shifted left are `spin-down' (red), while those along the parabola shifted right are `spin-up' (blue), as expected.\cite{Hoesch2004}

In  contrast, when the light is $s$-polarized, the photoelectron spin polarization ($P_y$) for Bi$_2$Se$_3$ is fully reversed (compare Fig.2~(c) and (d)), opposite to the expected surface state electron spin texture.
This is not the case for Au(111), where the spin polarization of photoelectrons is independent of the light polarization (compare Figs.~2(g) and (h)), showing that the effects seen in Bi$_2$Se$_3$ are not generic or trivial experimental artifacts.
This is true even for momentum points along both $k_x$ and $k_y$, measuring spin polarization along the $y$- and $z$-axes, and with both linear and circular light (see Supplementary Information).

\begin{figure*}\includegraphics[width=14cm]{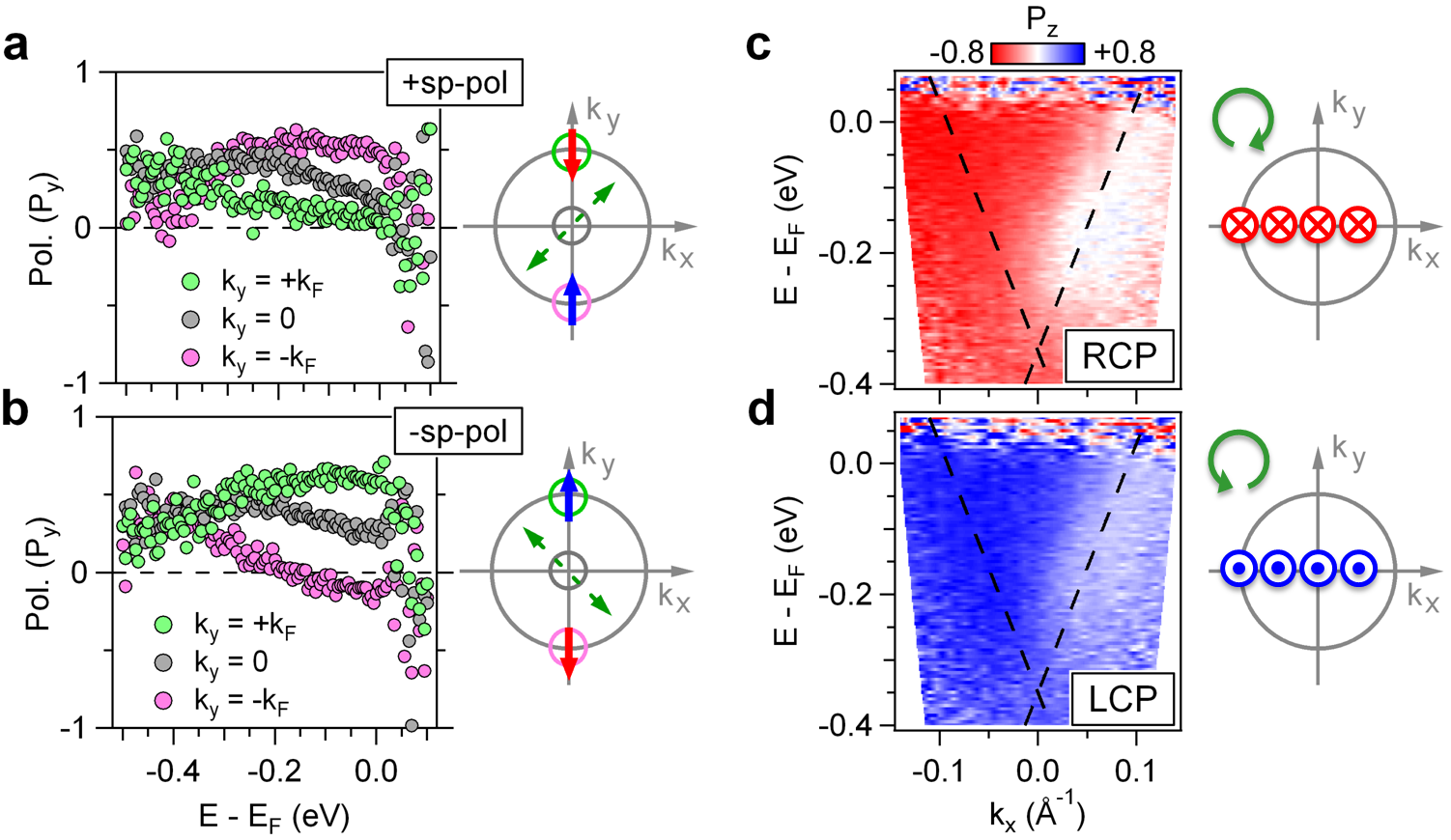}
\caption{\label{fig:SDb} \textbf{Bi$_2$Se$_3$ photoelectron spin polarizations with $\pm$sp-polarized and circularly polarized light.}
(a) Photoelectron $P_y$ curves at three values of $k_y$ along the $k_y$-axis, marked by small, color-coded circles in insets, for $+sp$-polarized light, whose $\hat{\epsilon}$ projections in the surface plane are shown by the green arrows in the insets.
(b) Same as (a), but for $-sp$-polarized light.
(c) Photoelectron $P_z$ maps as a function of binding energy and momentum along the $k_x$ axis, with right-hand circularly polarized light.  The dashed lines are guides to the eye, marking the topological surface state dispersion.
(d) Same as (c), but with left-hand circularly polarized light.}
\end{figure*}

The observations of strong dependence of the photoelectron spin polarization on photon polarization in Bi$_2$Se$_3$ demanded further investigation.
Figures~3(a,b) show photoelectron spin polarization ($P_y$) curves taken with `$\pm$$sp$-polarized' light, corresponding to the photon polarization vector being rotated to $\alpha_0=\pm 45^{\circ}$ (see Fig.~1(e)), halfway between $p$- and $s$-polarizations.
Three $P_y$ curves are shown for each, corresponding to the three momentum locations along the $k_y$ axis shown in the inset diagrams.
The theoretical surface state electron spin texture predicts $P_y=0$ at momenta along the $k_y$ axis (i.e. $k_x=0$) as the helical surface electrons are spin-polarized perpendicular to their momentum.
As above, an overall  $\mathbf{k}$-independent positive shift in $P_y$ in the measurement is due to the particular experimental geometries in these cases (see Supplementary Information).
The additional strong $k_y$ dependence in the data reveals the presence of a large radial component of the polarization, oriented as shown by the red and blue arrows in the diagrams, which was absent in previous measurements with $p$-polarized light.\cite{Jozwiak2011}
Such a radial component of the photoelectron spin polarization differs from the expected surface state electron spin texture which is primarily tangential at every point around the Fermi surface contour.
It is also clear that the measured radial components reverse between $+sp$- and $-sp$-polarized light geometries, again demonstrating control of the photoelectron spin polarization through the photon polarization. 

This control extends to the out-of-plane dimension through the use of circularly polarized light as shown in Figs.~3(c,d).
Specifically, panel (c) shows a full map of photoelectron polarization, similar to Figs.~2(c,d), but now measuring the out-of-plane spin component, $P_z$, and taken with right-hand circularly polarized light (RCP).
Throughout the map, photoelectrons are primarily polarized with spins directed into the surface, reaching values of $P_z=-0.8$.
Panel~(d) is a corresponding $P_z$ map taken with left-hand circularly polarized light (LCP), showing a full reversal with photoelectrons dominantly polarized with spin directed out of the surface, reaching values of $P_z=+0.8$.

The results shown in Figs.~1-3 reveal the ability to fully manipulate the spin polarization of photoelectrons from a topological insulator through control of the light polarization, to an extent not previously observed in any system.
They also illustrate nonequivalence of photoelectron and surface state spins in a topological insulator, contrary to the usual assumption in spin-resolved photoemission work.
Indeed, the results in Figs.~1~and~2 illustrate an interesting case of photoemission being dominated by a spin-flip process, an effect not previously experimentally observed (see Supplementary Information).

\begin{figure*} \includegraphics[width=12cm]{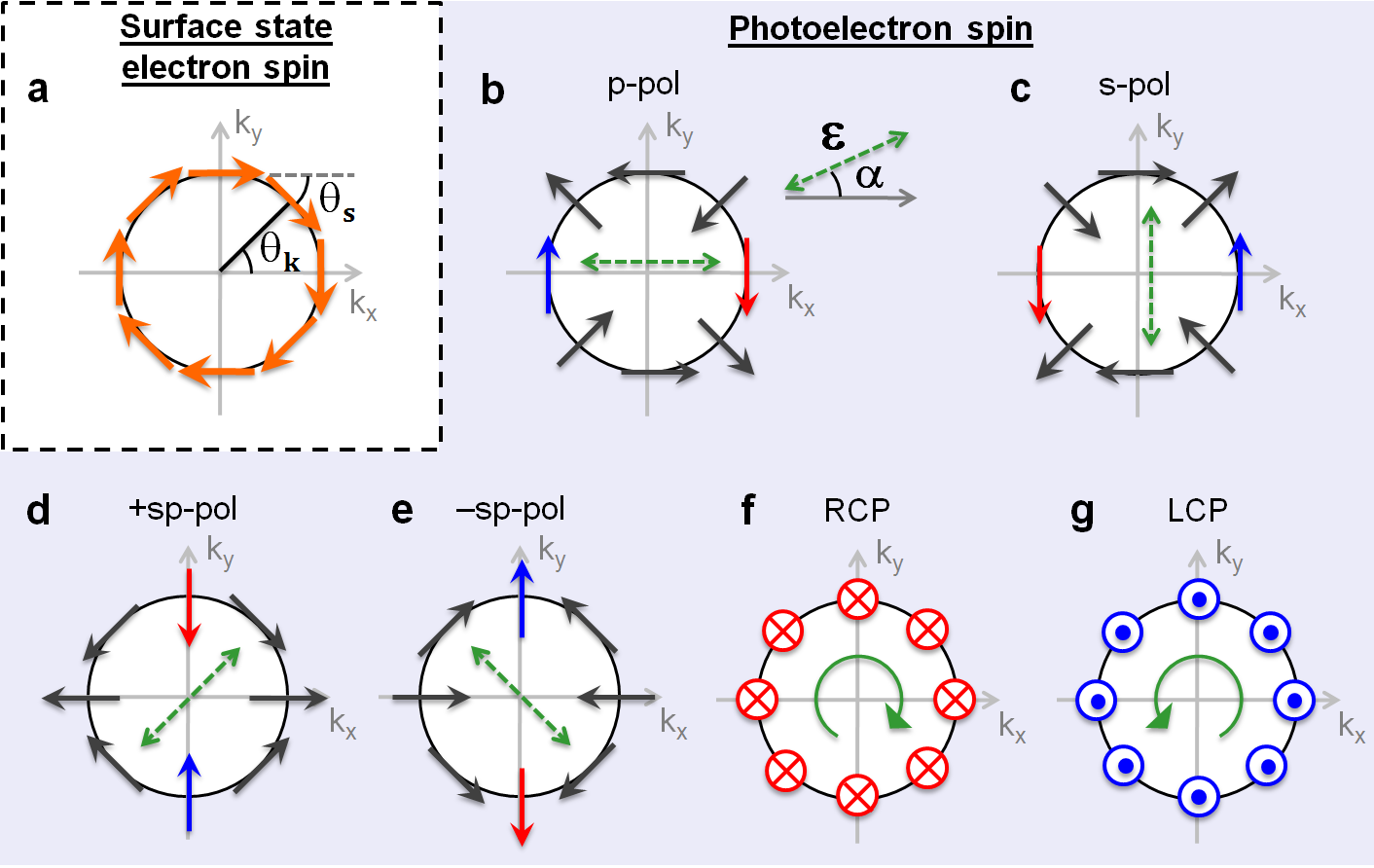}
\caption{\label{fig:SDa} \textbf{Calculated photoelectron spin textures from a topological insulator for various photon polarizations.}
(a) Spin texture of topological spin-helical Dirac electrons.  Arrows depict spin of surface state electrons, prior to photoemission.
(b-e) Calculated spin texture of photoelectrons from the same topological states, for various photon polarizations.\cite{CheolHwan}  Arrows depict the photoelectron spin polarization directions, using linearly polarized light.  The dashed green arrows mark the polarization vector, $\hat{\epsilon}$, projected onto the sample surface.  These correspond to `$p$-polarized' (b), `$s$-polarized' (c), and `$\pm sp$-polarized' light (d,e) in the current experiment.  The blue and red arrows correspond to the momentum positions and polarization directions consistent with the experimental data.
(f,g) Same as (b-e), but with normally incident circularly polarized photons.  Here, red crosses and blue dots depict photoelectron polarization into and out of the page along the $z$-axis.}
\end{figure*}

The primary aspects of our results are well described by considering the case of light incident normal to the Bi$_2$Se$_3$ surface, wherein the interaction Hamiltonian of the surface state electron and photon can be reduced to\cite{CheolHwan}
\begin{equation}
H_\textrm{int} \propto \left( \vec{\sigma} \times \hat{z} \right) \cdot \hat{\epsilon} ~ ,
\end{equation}
where $\vec{\sigma}$ is the spin Pauli matrix, $\hat{z}$ is the unit surface normal vector, and $\hat{\epsilon}$ is the photon polarization vector.
The presence of the spin Pauli operator readily shows that this interaction is capable of spin-flip transitions, counter to the usual assumption for such photoemission experiments.
Calculations of the corresponding spin dependent transition probabilities result in strong differences between predicted photoelectron spin polarization textures and the helical spin texture of the initial topological surface state,\cite{CheolHwan} as summarized in Fig.~4 for various photon polarizations, and in overall agreement with our measurements.

More specifically, the spin orientation of an electron in the helical surface state (Fig.~4(a)) can be expressed as 
\begin{equation}
\theta_\mathbf{s} = -\pi/2 + \theta_\mathbf{k} ~ , 
\end{equation}
where $\theta_\mathbf{s}$ is the angle between the $+x$ direction and the spin direction at momentum  $\mathbf{k}$, and $\theta_\mathbf{k}$ is the angle between the $+x$ direction and $\mathbf{k}$.
In the case of linearly polarized light, with $\hat{\epsilon}$ parallel to the sample surface, the corresponding photoelectrons become spin polarized along directions given by\cite{CheolHwan}
\begin{equation}
\theta_\mathbf{s}' = -\pi/2 + \left( 2\alpha - \theta_\mathbf{k} \right) ~ ,
\end{equation}
where $\alpha$ is the angle in the surface plane between the $+x$ direction and $\hat{\epsilon}$ (Fig.~4(b--e)).
Thus there is a difference between the initial spin state and photoelectron spin polarizations at all momenta except for where $\theta_\mathbf{k} = \alpha$.
Measurements within this typical geometry, such as Figs.~1(c)~and~2(c), have  $\theta_\mathbf{s}'=\theta_\mathbf{s}$.
This is likely why previous spin-resolved ARPES works did not find results counter to the expected surface state spin texture.

Equation~(3) well describes the observed reversal of the photoelectron $P_y$ from $p$-polarized ($\alpha=0$) to $s$-polarized ($\alpha=\pi/2$) light (see Figs.~1(c,d)~and~2(c,d)), and the general $\cos (2\alpha)$ dependence of the photoelectron $P_y$ measured at $\theta_\mathbf{k}=\pi$ (see Fig.~1(f)).
It also accounts for the large $P_y$ measurements of photoelectrons at $\theta_\mathbf{k}=\pi/2$~and~$3\pi/2$ measured with $\pm sp$-polarized light with $\alpha\sim\pm\pi/4$ (compare Figs.~3(a,b) with Figs.~4(d,e)). 
The above calculations also predict circularly polarized light to result in the out-of-plane directed photoelectron spin polarization textures shown in Fig.~4(f,g), in general agreement with the present data shown in Figs.~3(c,d) (see Supplementary Information).
The results in Figs.~3(c,d) are also in line with a recent theoretical study of photoemission from a related material, Bi$_2$Te$_3$.\cite{Mirhosseini2012}

The observed photon polarization dependent photoelectron spin flipping and spin texture control in Bi$_2$Se$_3$ thus stems in part from strong spin-orbit coupling in the material.
The observed absence of these effects for the Au(111) surface states (Fig.~2 and Supplementary Information), despite predictions of similar photoemission effects,\cite{Henk2003} may be due to weaker spin-orbit coupling and the resulting dominance of an additional Hamiltonian term due to the inversion symmetry breaking at the surface.\cite{Kim2012}

Independent of interpretation, our results demonstrate complete manipulation and control of photoelectron spin polarization from Bi$_2$Se$_3$.
This could be utilized in a variety of applications ranging from spintronics to photocathode sources of polarized electron beams.
In comparison to the widely used GaAs photocathode,\cite{Pierce1980} Bi$_2$Se$_3$ could provide larger polarizations and enhanced functionality with complete control of spin orientation in three dimensions.

Finally, we hope our findings will stimulate further studies of possible similar control of photoelectron spin in other materials.
Examples include other topological insulators, such as Bi$_{1-x}$Sb$_x$\cite{Hsieh2009} where the surface states extend to higher momenta to the Brillouin zone boundary where the dispersions are not linear.
Other Rashba systems where the splitting is much larger than in Au (e.g. two-dimensional electron gases on Bi$_2$Se$_3$\cite{King2011} or bulk BiTeI\cite{Ishizaka2011}) may provide insight, as well.


\ \

\section*{Methods}

Experiments were performed on Bi$_2$Se$_3$ single crystals grown by directional slow solidification in an inclined ampoule and cleaved \textit{in-situ} along the (111) plane in vacuum of $5\times 10^{-11}$ torr.
The Au(111) surface was prepared by \textit{in-situ} evaporation on a clean W(110) substrate according to standard methods.
High resolution spin-integrated ARPES data (Fig.~1(b,c)) were taken at beamline 4.0.3 at the Advanced Light Source with 35 eV linearly p-polarized photons, at a sample temperature of $<20$K.
The energy and momentum resolutions were $\sim$ 13 meV and 0.005 \AA$^{-1}$, respectively.
Spin-resolved ARPES data were taken with 5.99 eV laser light and a high-efficiency spin-resolved spectrometer utilizing time-of-flight (TOF) technique and low-energy exchange-scattering techniques.\cite{Jozwiak2010}
These data were taken at a sample temperature of $\sim$ 80 K, with instrumental energy and momentum resolutions of $\sim$ 15 meV and  0.02 \AA$^{-1}$, respectively.
The spectrometer acquires data as a function of binding energy in parallel, allowing high resolution full energy distribution curves (EDCs) to be acquired in 2-3 minutes, as opposed to several hours with conventional spin-resolved ARPES systems, thus precluding any surface aging effects (e.g. vacuum or laser exposure) during acquisition and enabling the wide coverage of experimental parameter space in the experiment.
Full two-dimensional energy-momentum polarization maps (Figs.~3(b,c)~and~4(c,d,g,h) are made up of 20-30 individual EDCs.
Each pair of maps (e.g. Fig.~4(c,d)) is taken simultaneously, alternating photon polarization after each EDC, such that photon polarization dependence in a pair of maps cannot be due to surface aging.
The momentum, or $\mathbf{k}$-vector probed in an EDC is scanned by rotating the crystal about the $y$ or $x$ axes, while the photon beam, photoelectron collection angle, and spin analysis axis are all held fixed, as shown in Fig.~1(f).

\begin{acknowledgments}
We thank G. Lebedev and W. Wan for work with the electron optics, and W. Zhang, D. A. Siegel, C. L. Smallwood, and T. Miller for useful discussions, H. Wang and R. A. Kaindl for advice with optics, and A. Bostwick for help with software development.
This work was supported by the Director, Office of Science, Office of Basic Energy Sciences, Division of Materials Sciences and Engineering, of the U.S. Department of Energy under Contract No. DE-AC02-05CH11231 (Lawrence Berkeley National Laboratory).
Higher photon energy photoemission work was performed at the Advanced Light Source, Lawrence Berkeley National Laboratory, which is supported by the Director, Office of Science, Office of Basic Energy Sciences, of the U.S. Department of Energy under Contract No. DE-AC02-05CH11231.
\end{acknowledgments}

\section*{Author Contributions}
C.J. developed the experimental system.  C.J., C.-H.P., and C.H. devised the experiment.  C.J. and K.G. carried out the experiment.  C.J. analyzed the experimental data.  Calculations were performed by C.-H.P., S.G.L. and D.-H. L.  Samples were prepared by C.R.R. and R.J.B.  Synchrotron data was acquired by J.D.D., C.J., and K.G.  Z.H. and A.L. were responsible for experiment planning and infrastructure.  All authors contributed to the interpretation and writing of the manuscript.

\section*{Additional information}
The authors declare no competing financial interests.
Supplementary Information accompanies this paper.


%% file: Jozwiaksi.tex
\begin{center}
\Large{\bf{Photoelectron spin-flipping and texture manipulation in a topological insulator}}
\large

\ \

Chris Jozwiak, Cheol-Hwan Park, Kenneth Gotlieb, Choongyu Hwang, Dung-Hai Lee,
 
Steven G. Louie, Jonathan D. Denlinger, Costel R. Rotundu, Robert J. Birgeneau, 

Zahid Hussain, and Alessandra Lanzara

\ \

\ \

{\bf{\Large{Supplemental Information:}}}

\end{center}

\ \

\ \

\begin{spacing}{1.5}

\section{Experimental setup}

The spin-integrated data of Fig.~1(a,b) was taken at the Merlin beamline, BL4.0.3, of the Advanced Light Source (Berkeley, CA, USA), using a commercial Scienta R8000 hemispherical analyzer with a photon energy $h\nu=35$~ eV, and linear $p$-polarization.
All other data, both spin-resolved and spin-integrated, were taken with the `spin-TOF' spectrometer with a laser-based light source in the geometry shown in Fig.~S1, as discussed below.

The laser source is a cavity-dumped, mode-locked Ti:Sapphire oscillator pumped by a 6 W frequency doubled Nd:YVO$_4$ laser.
This oscillator generates $\sim$ 150 fs pulses, tuned to 828 nm at a repetition rate of 54.3/$n$~MHz, where $n$ is an integer.
For the present work, $n$ was set to 10, for a repetition rate of $\sim$ 5 MHz -- this approaches the maximum repetition rate compatible with the `spin-TOF' spectrometer in typical conditions.
The oscillator output is frequency-quadrupled through cascaded, type-I phase-matched second harmonic generation in two beta barium borate (BBO) crystals of 2 and 5~mm thicknesses, producing a 207 nm (5.99 eV) beam, with a measured bandwidth of $\sim$~4~meV and pulse lengths estimated to be several picoseconds.
The polarization of the 6 eV beam is straightforwardly controlled with zero-order half-wave and quarter-wave plates, providing linearly and circularly polarized light, respectively.
The beam is then focused into the main vacuum chamber through a UV-grade fused silica viewport and onto the sample surface.

The experimental geometry is schematically shown in Fig.~S1.
The $x$, $y$, and $z$ axes reference a fixed coordinate system in the lab, with the origin located at the simultaneous intersection of the sample surface, the photon beam, and the electron-optical axis of the photoelectron spectrometer.
The photon beam is incident in the $xz$ plane, shaded gray in the figure, at a fixed 45$^{\circ}$ angle from the $x$ axis.
The photon beam, when linearly polarized, can have its polarization vector oriented at any angle $\alpha_0$ between $p$- and $s$-polarization geometries, as shown.
The photon polarization vector is aligned within the $xz$ plane for $p$-polarization and is along the $y$ axis for $s$-polarization, respectively, as defined in the present manuscript.
Circular polarizations of either helicity can also be selected, with right-hand circularly polarized (RCP) light defined in the figure.

Photoelectrons emitted along the $z$ axis are collected by the spectrometer, with an angular acceptance of $\sim \pm$1$^{\circ}$.
This translates to a momentum resolution of $\sim$ 0.02 \AA$^{-1}$.
Selecting the momentum to be probed requires rotation of the sample surface with respect to the fixed spectrometer.
With the sample surface aligned to the fixed $xy$ plane as drawn in Fig.~S1, emission at $\Gamma$, or $(k_x,k_y) = (0,0)$, is probed.
The value of $k_x$ is scanned by rotating the sample about the $y$ axis (the $\theta$ rotation in Fig.~S1), while the value of $k_y$ is scanned by rotating the sample about the $x$ axis (the $\beta$ rotation in Fig.~S1).

\begin{figure*} \includegraphics[width=14cm]{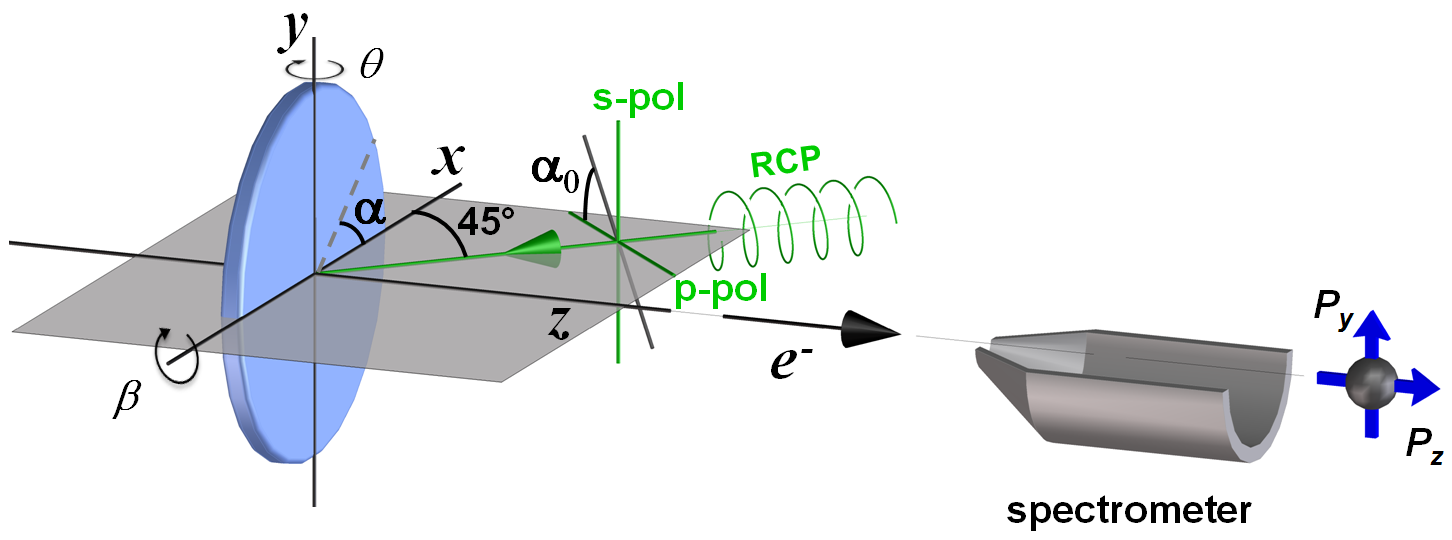}
\caption{\label{fig:geo} \textbf{Experimental geometry.}
Schematic diagram of the experimental geometry.  The $x$, $y$, and $z$ axes reference a fixed coordinate system.  The photon beam is incident within the $xz$-plane, at a fixed angle from the $x$ axis.  Photoelectrons emitted along the fixed $z$ axis, shown by the black arrow, are collected by the spectrometer, which is sensitive to spin along the $y$ and $z$ axes.  The photons can be linearly polarized with any orientation between $p$- and $s$-polarizations, defined by the angle $\alpha_0$.  The photons can also be circularly polarized, with either helicity.  Right-hand circularly polarized light (RCP) is shown. 
}
\end{figure*}

The `spin-TOF' spectrometer is a custom built spin-resolved photoelectron spectrometer for high efficiency acquisition of spin-resolved photoelectron energy distribution curves (EDCs) with high energy and angular resolution.
It is described in detail in Ref.~\onlinecite{Jozwiak2010}.
Photoelectron kinetic energies are resolved in parallel through the `time-of-flight' (TOF) technique, in which the total photoelectron transit time from emission to detection is accurately measured, providing an energy resolution of $\sim$~15~meV.
This technique requires the light source to be pulsed, and requires an adequate timing window between pulses, thus setting an upper limit on the laser repetition rate.
In the present case, this limit is $\sim$ 10 MHz in standard conditions.
Spin is resolved through differential measurement of the relative reflectivity of the photoelectrons scattered from the surface of a magnetic thin film.
This is described in detail in Ref.~\onlinecite{Jozwiak2010} and references therein.
Due to specific details of the scattering geometry used in the `spin-TOF' spectrometer, spin can be resolved along the fixed $y$ or $z$ axes, providing measurements of $P_y$ and $P_z$, as depicted in Fig.~S1.
The directions of positive $P_y$ and $P_z$ as used in the manuscript are defined by the directions of the corresponding arrows.

The overall efficiency of the spectrometer and laser allowed a high acquisition speed.
This efficiency was critical for the current study which included spin-resolved data through a wide parameter space -- spin was resolved along two axes (i.e. $P_y$ and $P_z$) and measured as a function of binding energy, momentum, and many photon polarizations.
A high level of statistics was achieved for a single spin-resolved EDC in only 2-3 minutes, allowing spin resolved data to be taken with multiple photon polarizations in rapid succession without the sample surface degrading or altering due to finite residual vacuum conditions (the vacuum chamber pressure was $\sim 6\times10^{-11}$~torr).
Full spin-resolved maps were quickly acquired in $\sim$~1~hour, by measuring $\sim$~ 30 successive spin-resolved EDCs while scanning momentum (emission angle) in small steps.
When acquiring such spin-resolved maps with different photon polarizations for direct comparison (i.e. Fig.~2(c,d,g,h) and Fig.~3(c,d)), the maps were acquired `interlaced', with an EDC at one $k$ (emission angle) being successively taken with each photon polarization before moving to the next $k$ (angle).
This approach provides a better direct measure of photon polarization dependence, free of any time dependent effects that may be introduced if the maps were acquired separately, one after the other.

\section{Extrinsic spin-polarization effects induced by spin-orbit coupling}

Although often forgotten, it is known that spin polarization effects can occur in photoemission due to spin-orbit coupling, causing the spin polarization of the photoelectrons to be different from that of the corresponding initial states.
This is most easily exemplified in cases where spin-polarized photoelectrons are measured from unpolarized initial states, such as unpolarized atoms or from spin-degenerate states in non-magnetic solids.
The `Fano effect', in which photoelectrons from the $s$ orbitals of unpolarized atoms can be spin-polarized using circular polarized light,\cite{Fano1969} is such an example.
Less intuitively, it was also shown that photoelectrons emitted from spin-degenerate atomic subshells of orbital angular momentum $l > 0$ into well defined angular directions can also be spin-polarized, even when using linear and unpolarized light.\cite{Lee1974,Cherepkov1979}

It should be noted that in these cases, the photoemission dipole operator considered does not actually change the orientation of the electron spin through the photoemission process.\cite{Cherepkov1979}
Instead, the measured spin polarization results from effective spin-dependent photoemission matrix elements (SMEs) which effectively lead to selective emission of electrons with a particular spin orientation.
As an example, the SME-induced photoelectron polarization vector for photoionization of atoms with linearly polarized light is\cite{Cherepkov1979}
\begin{equation}\label{eqn:psme}
\boldsymbol{\vec P}_{\mathrm{SME}} = \frac{2\xi \left( \boldsymbol{\hat{k}}_e \cdot \boldsymbol{\hat{\epsilon}} \right) }{1+\beta\left(\frac{3}{2}(\boldsymbol{\hat{k}}_e \cdot \boldsymbol{\hat{\epsilon}})^2 - \frac{1}{2}\right)}\left[ \boldsymbol{\hat{k}}_e \times \boldsymbol{\hat{\epsilon}} \right] \,,
\end{equation}
where $\boldsymbol{\hat{k}}_e$ and $\boldsymbol{\hat{\epsilon}}$ are the outgoing photoelectron and photon polarization unit vectors, respectively.
The denominator of Eqn.~(\ref{eqn:psme}) is due to the angular distribution of photoemission where $\beta$ is the asymmetry parameter.
The parameter $\xi$ reflects the interference between the possible $l+1$ and $l-1$ continuum photoelectron states and is the source of the spin dependence.
Thus, this SME-induced polarization is due to the spin-orbit interaction.
Like a great many other matrix element related effects in ARPES, $\boldsymbol{\vec P}_{\mathrm{SME}}$ is dependent on details of the initial and final photoelectron states, and therefore also photon energy.

Equation~(\ref{eqn:psme}) also shows that both the magnitude and orientation of $\boldsymbol{\vec P}_\mathrm{SME}$ are dependent on the orientations of the photon polarization and the outgoing photoelectrons.
The geometrical terms in Eqn.~(\ref{eqn:psme}) ($\boldsymbol{\hat{k}}_e \cdot \boldsymbol{\hat{\epsilon}}$ and $\boldsymbol{\hat{k}}_e \times \boldsymbol{\hat{\epsilon}}$) are required by symmetry: parity conservation requires $\boldsymbol{\vec P}_\mathrm{SME}$ to be perpendicular to the reaction plane formed by $\boldsymbol{\hat{\epsilon}}$ and $\boldsymbol{\hat{k}}_e$, or more generally to any mirror planes of the complete system.\cite{Kesslerbook}
For the case of circularly polarized light, the corresponding equation is more complicated, involving a component of $\boldsymbol{\vec P}_\mathrm{SME}$ perpendicular to the reaction plane formed by $\boldsymbol{\hat{k}}_e$ and the propagation vector of the photon flux, similar to above, and an additional component along the propagation vector of the photon flux that changes sign with the photon helicity.\cite{Cherepkov1979}

In addition to being observed in atomic photoionization,\cite{Heinzmann1979} spin polarized photoemission qualitatively described by Eqn.~(\ref{eqn:psme}) has been observed in solid-state photoemission from core levels of nonmagnetic systems, including the Cu $2p$ and $3p$ (Ref.~\onlinecite{Roth1994}), W $4f$ (Ref.~\onlinecite{Rose1996}), and Pt $4d$ and $4f$ levels (Ref.~\onlinecite{Yu2008}), as well as the Bi $5d$ levels in Bi$_2$Se$_3$.\cite{Jozwiak2011}
Similar SMEs have been both predicted\cite{Tamura1987,Tamura1991,Tamura1991a,Henk1994} and observed\cite{Schmiedeskamp1988,Schmiedeskamp1991,Irmer1992,Irmer1994,Irmer1995,Irmer1996} in various forms in the valence bands of Pt and Au single crystals.

We have previously shown such SMEs induce spin polarized photoemission from the bulk valence and conduction bands of Bi$_2$Se$_3$,\cite{Jozwiak2011} as well, which can significantly impact on the spin-resolved ARPES data from this system.
Figure~S2(a) shows an ARPES intensity map of Bi$_2$Se$_3$.
Panel (b) shows the $y$ component of the measured photoelectron spin polarization curves, corresponding to the vertical line cuts marked in (a).
Three curves are shown (corresponding to $k_x=0,\pm k_\textrm{F}$) at three different photon energies.
Each data set in Fig.~S2(b) is taken in the geometry of Fig.~S1 with linear, $p$-polarized light.
In the absence of SME-related effects, one expects to observe large values of $P_y$ due to the spin polarization of the helical spin texture of the topological surface state.
This texture dictates a strongly $k_x$-dependent $P_y$, with $P_y$ at $k_x=+ k_\textrm{F}$ to be reversed from $P_y$ at $k_x=- k_\textrm{F}$.
Following the symmetry requirements discussed above, any SME-induced spin polarization must be oriented along the $y$ axis, and is largely determined by the angle between the photoelectron emission direction and the photon polarization vector (from Eqn.~(\ref{eqn:psme}), which is qualitatively applicable here).
As this angle stays fixed in the experiment while $k$ is scanned, any SME-induced spin polarization is effectively independent of $k$.

\begin{figure*} \includegraphics[width=16cm]{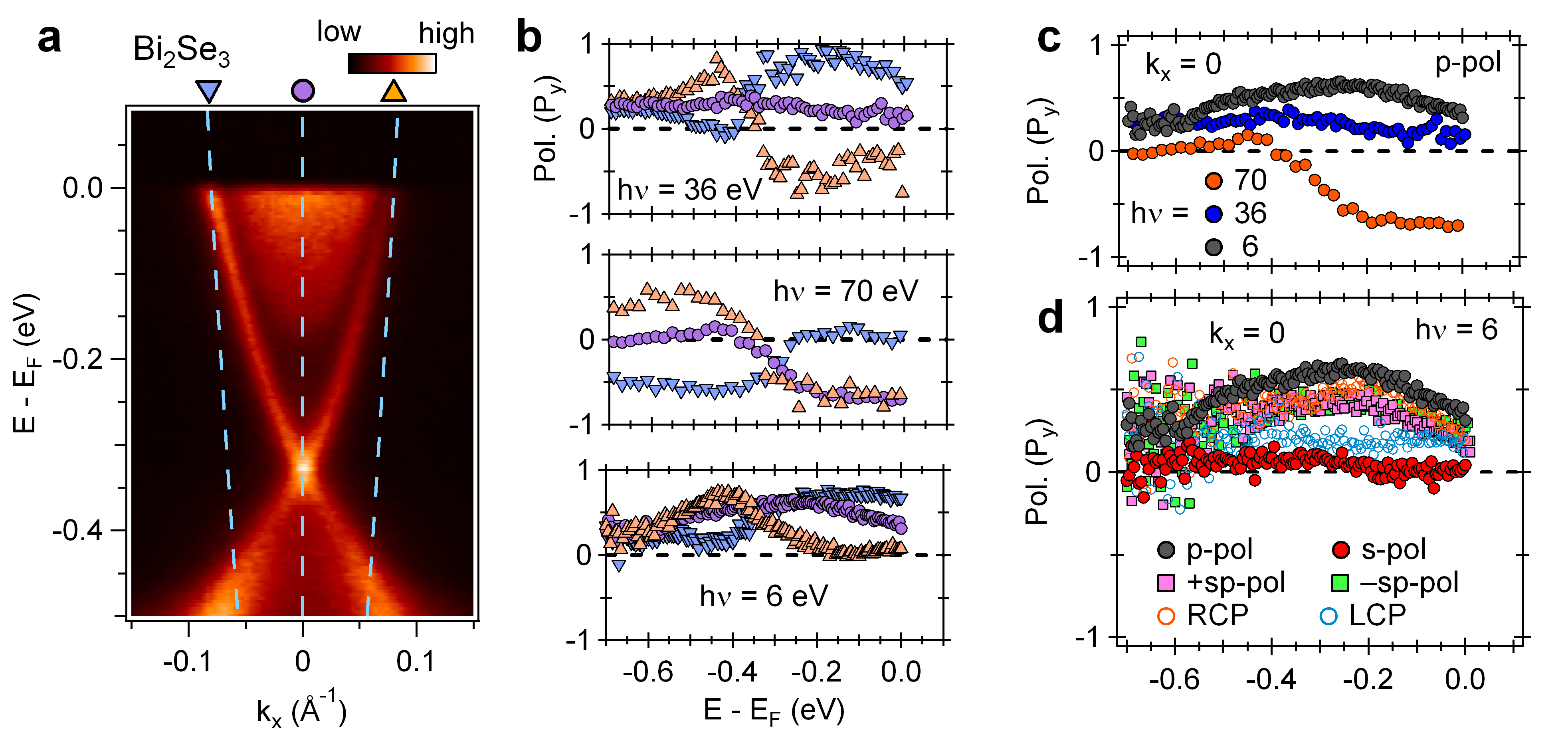}
\caption{\label{fig:fig1} \textbf{The dependence of photoelectron background spin on photon energy and polarization.}  
(a) ARPES intensity map of Bi$_2$Se$_3$ as a function of binding energy and momentum, along $\Gamma$M.  Taken with $h\nu=35$ eV.
(b) The measured $y$ component of the photoelectron spin polarization, $P_y$, as a function of binding energy at a given momentum.  Each panel contains curves corresponding to the momenta of the vertical cuts shown in (a), labeled by marker.  Each panel corresponds to data taken with the specified photon energy, taken with the $p$-polarized light geometry.
(c) Direct comparison of the photoelectron $P_y$ at $\Gamma$ ($k_x=0$) from (b) at each photon energy. 
(d) The photoelectron $P_y$ at $\Gamma$, measured with the laser ($h\nu=6$~eV), with various photon polarizations.
}
\end{figure*}

Indeed, the $P_y$ curves in Fig.~S2(b) show both $k_x$-dependent and $k_x$-independent contributions.
As the $k_x$-dependent contribution should have $P_y=0$ at $k_x=0$, the measured $P_y$ curves at $k_x=0$ (purple circles) can be taken as a measure of the $k_x$-independent contribution caused by SMEs.
In the top panel, taken with $h\nu=36$~eV photons, the $k_x=0$ curve is non-zero at all binding energies, reflective of SMEs throughout the electronic structure, including the bulk.
The $k$-dependent component of $P_y$, which is reflective of the surface state spin texture, can be seen in addition to this: the $P_y$ curve at $k_x=+k_\textrm{F}$ is the `inverse' of that at $k_x=-k_\textrm{F}$, approximately inverted about the $k_x=0$ curve.
These two curves each exhibit two peaks, a maximum and minimum, resulting from the opposite spin textures of the upper and lower halves of the Dirac cone.
These features are discussed in detail in Ref.~\onlinecite{Jozwiak2011}.
The $h\nu=36$~and~70~eV data were taken previously with synchrotron light\cite{Jozwiak2011}, while the $h\nu=6$~eV data were taken with the laser source as part of the current experiment.

The three panels of Fig.~S2(b) show that while the characteristic $k$-dependent component of the $P_y$ curves remains at each photon energy, the $k$-independent SME-induced component is strongly photon energy dependent -- it even changes sign between $h\nu=36$~and~70~eV.
To illustrate this more clearly, the curves at $k_x=0$ in each panel of Fig.~S2(b) are plotted again for direct comparison in Fig.~S2(c).
Furthermore, the SME-induced photoelectron spin polarization is strongly dependent on the photon polarization.
Figure~S2(d) shows the same $P_y$ curves at $k_x=0$ measured with the 6~eV laser at several photon polarization geometries.
As in Eqn.~(\ref{eqn:psme}), the SME-induced photoelectron spin polarization is maximum for $p$-polarized light ($\boldsymbol{\hat{k}}_e \cdot \boldsymbol{\hat{\epsilon}}$ is maximum), and is very close to zero for $s$-polarized light ($\boldsymbol{\hat{k}}_e \cdot \boldsymbol{\hat{\epsilon}}=0$).
 
Thus, the photoelectron $P_y$ in Bi$_2$Se$_3$ measured with the $h\nu=6$~eV laser in the current geometry exhibits a strong SME-induced component, particularly with $p$-polarized light, which is $k$-independent and distinct from the intrinsic surface state spin texture related to the topological ordering.
This effectively induces a loosely qualitative `offset' or `shift' to the measured photoelectron $P_y$ curves, visible in each panel in Fig.~S2(b).
This `shift' is then also visible in corresponding full $P_y$ maps.
For instance, Fig.~S3(a) shows a map of $P_y$ as a function of binding energy and $k_x$.
This false color scale image is the result of taking $\sim$ 30 individual $P_y$ curves such as those shown in Fig.~S2(b), as a function of emission angle, and mapping the data to energy-momentum space.
The map exhibits the momentum-dependent component of $P_y$ that reflects the underlying helical spin texture of the surface state: on the left there is a dark blue (very positive $P_y$) streak following the surface state dispersion, and on the right there is a slightly red (slightly negative $P_y$) streak following the other side of the surface state dispersion.
However, since the data was taken with $p$-polarized light, there is a qualitative `shift' to positive  $P_y$ (or `blue' in the figure).
For instance, signal in between the dispersions (near $E_\textrm{F}$ and $k_x=0$), due to the bulk conduction band which is assumed spin-degenerate in the crystal, appears blue (moderately positive $P_y$).

\begin{figure*} \includegraphics[width=12cm]{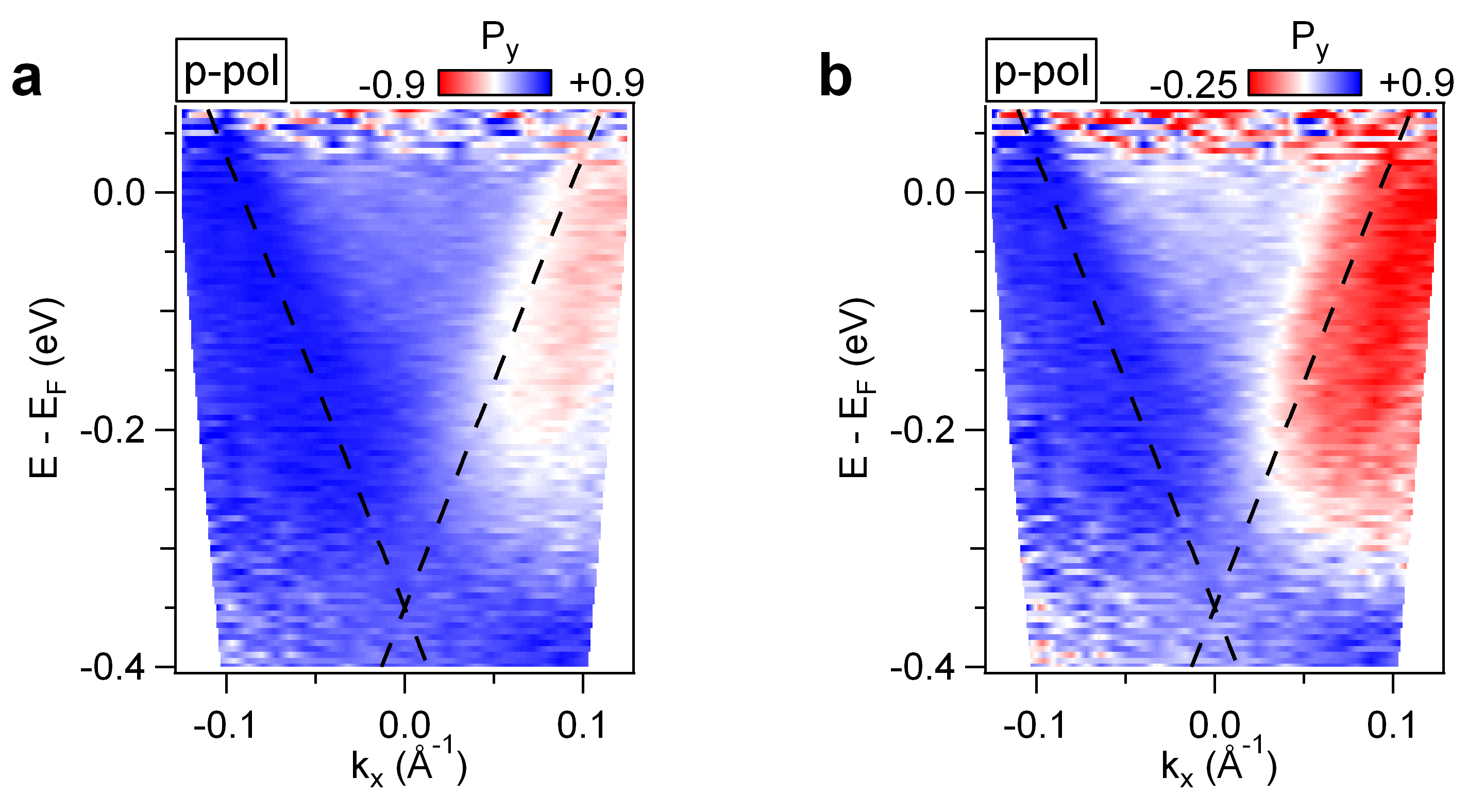}
\caption{\label{fig:fig1} \textbf{Asymmetric color scale to counter induced polarization asymmetry.}  
(a) Photoelectron $P_y$ map as a function of binding energy and momentum along the $k_x$ axis ($\Gamma$M), measured with $p$-polarized photons with $h\nu=6$~eV.  Here the color scale is symmetric, ranging from $P_y = -0.9$~to~+0.9, with `white' corresponding to $P_y=0$.  The dashed lines are guides to the eye, marking the topological surface state dispersion.
(b) Same as (a), but with an asymmetric color scale ranging from $P_y=-0.25$~to~+0.9.
}
\end{figure*}

This SME-induced `shift' can be qualitatively removed simply by displaying the map with an asymmetric color scale, such that the color `white' approximately corresponds to the SME-induced component.
Figure~S3(b) shows the same data as panel (a), but with an asymmetric color scale, resulting in a map that more readily displays the $P_y$ texture related to that of the helical Dirac surface state.
This map is the one displayed in Fig.~2(c) of the main paper, as the $k$-independent SME-induced effects are not the primary focus of the current work.
The similar map in Fig.~2(d) of the main paper was acquired with $s$-polarized light, and so the SME-induced $P_y$ goes to zero, as discussed above.
Thus there is no `shift' to the map, and it displays the $k$-dependent $P_y$ quite well with a symmetric color scale.
While this $s$-polarized light geometry is free from SME-induced effects, the photoelectron $P_y$ texture is opposite to the intrinsic helical spin texture of the surface state electrons, as discussed in the main text.
With the present geometry and photon energy (6 eV), the Au(111) sample appears free of SME-induced $P_y$ even with the $p$-polarized light, and thus the $P_y$ maps of Fig.~2(g)~and~(h) are both shown with symmetric color scales.

This SME-induced `shift' in $P_y$ is also present for $\pm sp$-polarized light geometries, used in Fig.~3(a,b) of the main paper, although quantitatively less than for $p$-polarized light.
The SME effect on $P_y$ for both $+sp$- and $-sp$-polarizations should be the same, following Eqn.~(\ref{eqn:psme}); our measurement is in agreement with this reasoning (e.g. see Fig. S2(d)).
Just as for the above discussion with $p$-polarized light, this effect `shifts' the $P_y$ curves of Figs.~3(a,b) slightly to more positive values.
On top of this, there remains a clear $k_y$ dependence consistent with a radial component of the photoelectron spin texture shown in the inset and discussed in the text.

\section{Novel spin manipulation in photoemission}

As discussed above, there are numerous observed and understood cases of photoemission inducing a spin-polarization in the photoelectrons that is different from the spin-polarization of the initial state crystal electrons.
These effects, however, are normally understood as being due to spin-dependent transition matrix elements of spin-conserving transitions, which preferentially excite electrons of a particular spin, resulting in photoelectron ensembles with altered spin polarization.
The cases where these effects have been observed  involved measuring slight spin polarizations in photoelectrons corresponding to unpolarized initial states.\cite{Kessler1970,Heinzmann1979,Roth1994,Rose1996,Yu2008,Starke1996,Schmiedeskamp1988,Schmiedeskamp1991,Irmer1992,Irmer1994,Irmer1995,Irmer1996} 

The present observations in Bi$_2$Se$_3$ of photoelectron spin polarization flipping dependent on photon polarization are unique.
In this case, the initial state crystal electrons are in fact highly spin polarized to begin with and are emitted with different or even opposite spin polarization, dependent on photon polarization, suggesting a fundamental difference from previous spin effects seen in photoemission.
This is an interesting demonstration of low energy photoemission being dominated by a spin-flip transition, which are usually assumed to contribute negligibly compared to spin-conserving transitions.\cite{Feuchtwang1978a,Federbook}
To our knowledge, the present results are the first observation of photoelectron spin flipping dependent on linear photon polarization and such a wide extent of demonstrated control of photoelectron spin in three dimensions through manipulation of only the photon polarization.

\section{Impact of $A_z$ component of incident light}

Reference~\onlinecite{CheolHwan} presents the interaction Hamiltonian of the surface state electron and photon as
\begin{equation}\label{eqn:fullHam}
H_\textrm{int} \propto \left(A_y\sigma_x - A_x\sigma_y \right) + i\gamma A_z I ~ ,
\end{equation}
where $\mathbf{A}$ is the vector potential of the photon field, $\sigma_x$ and $\sigma_y$ are the Pauli matrices, and $I$ is the $2\times 2$ identity matrix.
This expression corresponds to equation (18) of Ref.~\onlinecite{CheolHwan}, ignoring an overall coefficient and using $\gamma$ for $2\beta/\alpha$.
In this expression, the $x,y,z$ components reference the sample coordinate system where $x$ and $y$ are the horizontal and vertical axes in the sample plane, and $z$ is the axis along the sample surface normal.

As discussed in Ref.~\onlinecite{CheolHwan}, since the last term above is proportional to the identity matrix, any $A_z$ component (i.e. any out-of-surface-plane component) of the photon polarization contributes to spin-conserving photoemission and cannot alone alter the spin polarization of photoemitted electrons.
Although the present experiment contains a finite $A_z$ component for all photon polarizations except for $s$-polarized light (see Fig.~S1), it is ignored in the main text in order to focus on the new physics contained in the spin-flip terms proportional to $A_y$ and $A_x$.
Specifically, in the main text the interaction Hamiltonian of the surface state electron and photon is expressed in equation (1) of the main text as
\begin{equation}\label{eqn:shortHam}
H_\textrm{int} \propto \left( \vec{\sigma} \times \hat{z} \right) \cdot \hat{\epsilon} ~ ,
\end{equation}
where $\vec{\sigma}$ is the spin Pauli matrix, $\hat{z}$ is the unit surface normal vector, and $\hat{\epsilon}$ is the photon polarization vector.
This expression is equivalent to Eqn.~(\ref{eqn:fullHam}) assuming $A_z=0$.
As presented in the main text, this assumption leads to the simple picture presented in Fig.~4 of the main text, which has good overall agreement with the measured photoelectron spin polarizations and is consistent with the observed photon polarization dependence.

For more insight, however, it is worthwhile to also consider the impact of the $A_z$ component.
Since to first order it contributes to spin-conserving photoemission, its impact is expected to decrease the effect of the spin-flipping photoemission, and thus decrease the photon polarization dependence of the photoelectron spin polarization.
For example, the magnitude of the photon polarization dependence shown in Fig.~3(a,b) of the main text is smaller than would be expected from Eqn.~(\ref{eqn:fullHam}) assuming $A_z = 0$, consistent with the finite $A_z$ present in the experiment reducing the photon polarization dependence.

\begin{figure*} \includegraphics[width=10cm]{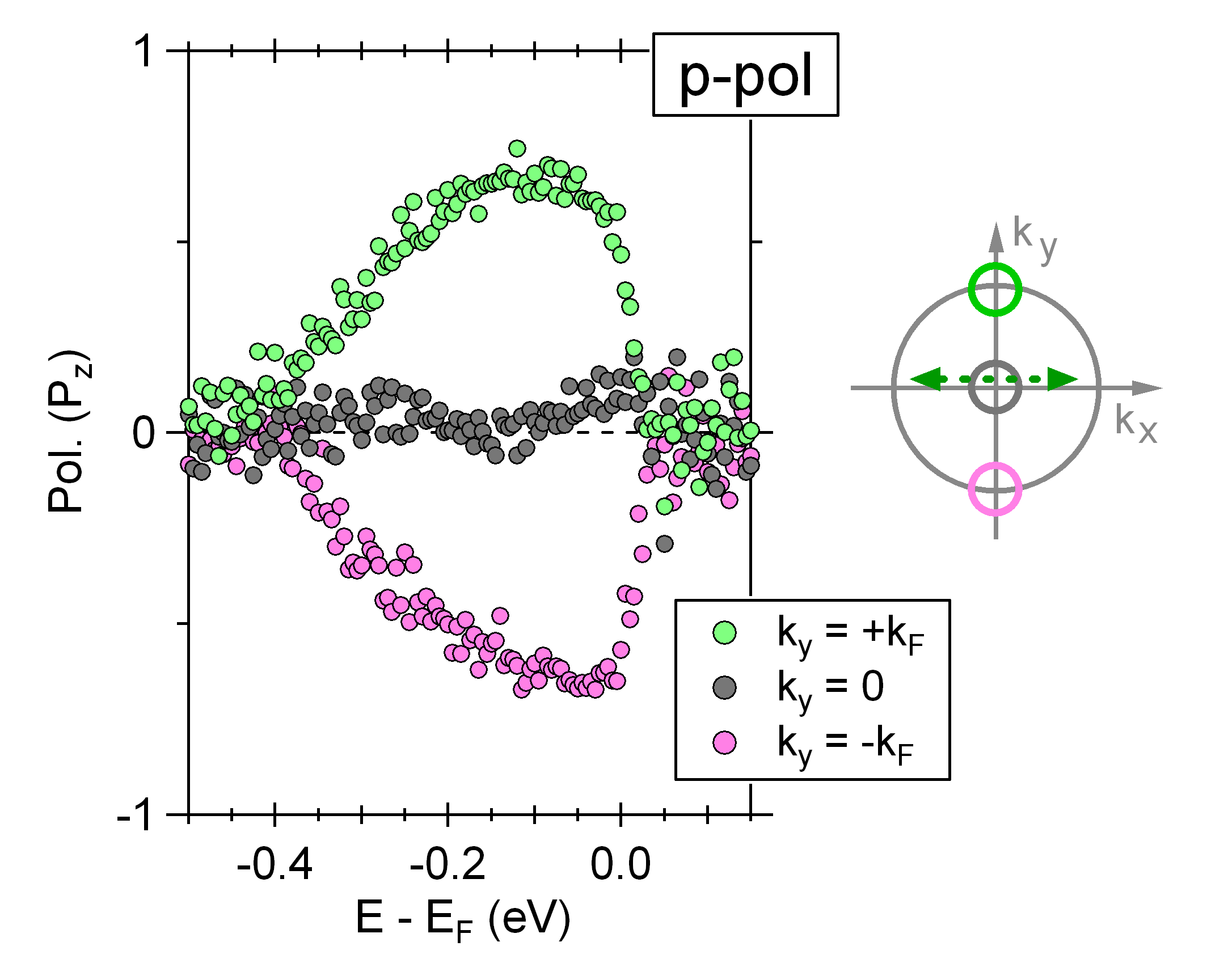}
\caption{\label{fig:fig1} \textbf{Large photoelectron $P_z$ induced by off-normal incidence of light.}  
Measured photoelectron $P_z$ curves as a function of binding energy at $(k_x,k_y)=(0,-k_F), (0,0), \textrm{and}, (0,+k_F)$, corresponding to the locations shown by the color-coded circles in the Fermi surface diagram.  The light was $p$-polarized, with the photon polarization component in the sample surface plane along the $k_x$ direction, as shown by the green arrow in the inset.  The photon polarization also has a significant component along the direction normal to the sample surface plane.
}
\end{figure*}

Perhaps more interestingly,  there should also be effects due to interference of the spin-conserving $A_z$ term and the spin-flip terms in Eqn.~(\ref{eqn:fullHam}).
These effects should be easiest to observe in cases where the spin polarization due to the individual terms is small.
For instance, we consider the case of measuring the out-of-plane spin polarization component, or $P_z$, at momentum positions along the $k_y$ axis using $p$-polarized light.
As the Fermi surface of the Bi$_2$Se$_3$ sample is mostly isotropic, without much hexagonal warping which can introduce out-of-plane spin polarization in the surface state \cite{Fu2009}, the surface state spin is primarily in-plane,\cite{Jozwiak2011} following the helical spin texture.
Figure~S4 shows several $P_z$ measurements, taken with $p$-polarized light and at momenta along the $k_y$ axis at $k_y = -k_F,0,\textrm{and} +k_F$.
In contrast to the expected surface state spin texture, surprisingly high values of photoelectron $P_z$ are measured, reaching nearly $\pm70\%$ at $k_y=\pm k_F$, respectively.

The interaction Hamiltonian in Eqn.~(\ref{eqn:fullHam}) is qualitatively consistent with these results as follows.
Assuming only spin-conserving photoemission, the photoelectron spin is expected to be directed along the $\pm k_x$ direction at $k = \pm k_y$, and thus $P_z$ is expected to be zero.
Likewise, as presented in Ref.~\onlinecite{CheolHwan} and Fig.~4 of the main text, assuming only the spin-flip terms above, the photoelectron spin polarization is expected to oppositely directed along the $\pm k_x$ axis, and thus $P_z$ is still expected to be zero.
However, consideration of finite contributions of both terms predicts large values of $P_z$ with reversed signs at $k_y>0$ and $k_y<0$.
Specifically, calculations following Ref.~\onlinecite{CheolHwan} predict values of $P_z$ to exceed $\pm90\%$ at $k_y=\pm k_F$, respectively, for $\gamma^2=2.0$ and $\pm75\%$ at $k_y=\pm k_F$ for $\gamma^2=0.25$.
Note these quantitative predictions do not take into account scattering effects which can reduce measured photoelectron spin polarizations.
Note also that they do not take into account the extrinsic effects due to spin-orbit induced spin-dependent matrix elements (SMEs) discussed in Section SI II, which are also present due to the off-normal incident light.

Additional insights can likely be gained by further investigation of the photoelectron spin polarization (in three dimensions) in a large variety of momenta and photon polarization geometries.
Rigorous quantitative comparisons with calculations will require the SMEs, as well as other possible effects, to be taken into account.

The following Sections (SI V and SI VI) discuss particular aspects of the data presented in the main text which are likely affected by the finite $A_z$ component in the experiment.

\section{Fit to the $P_y$ dependence on photon polarization rotation}

Following the approaches presented in the main text and Ref.~\onlinecite{CheolHwan}, assuming the light is incident normal to the sample surface and the photon polarization vector is entirely within the surface plane, the $y$ component of the spin polarization ($P_y$) of photoelectrons emitted from the surface state of a topological insulator using linearly polarized light will be
\begin{equation}
P_y = \frac{n}{A_x^2 + A_y^2} \left[ \left( A_y^2 - A_x^2 \right) \cos{\theta_\mathbf{k}} - 2A_x A_y \sin{\theta_\mathbf{k}} \right] ~.
\end{equation}
In the above, $A_x$ and $A_y$ are the components of the photon polarization vector along the $x$ and $y$ axes, respectively, in the sample surface plane, $\theta_\mathbf{k}$ is the angle with respect to the $x$ axis of the in-plane momentum vector, $\mathbf{k}$, being probed, and $n=\pm1$ for the upper and lower cones of the Dirac dispersion, respectively.
For the upper band at $E_F$, and for $\theta_\mathbf{k}=\pi$ (corresponding to the measurement in Fig.~1(f) of the main paper), and the geometry of the present experiment, this gives,
\begin{equation}
P_y = \frac{\cos^2{\theta} \cos^2{\alpha_0} - \sin^2{\alpha_0}}{\cos^2{\theta} \cos^2{\alpha_0} + \sin^2{\alpha_0}},
\end{equation}
where $\alpha_0$ is defined in Fig.~1(e)~and~S1, and $\theta$ is the angle between the incident light propagation vector and the sample normal.
As drawn in Fig.~1(e)~and~S1, $\theta=45^{\circ}$, however the sample must be rotated about the $y$-axis for the measurement to reach this point in momentum space, such that for $\theta_\mathbf{k}=\pi$, $\theta=36^{\circ}$.

We found that the above equation does not fit the slight asymmetry of the data shown in of Fig.~1(f) perfectly, likely due to the finite component of the photon polarization along the out-of-surface-plane direction ($A_z$), which is largest for $\alpha_0=0$, and vanishes for $\alpha_0=90^{\circ}$.
According to Ref.~\onlinecite{CheolHwan}, the $A_z$ component can contribute to the total photoemission through purely spin-conserving emission.
We can add the influence of this contribution to the total measured spin polarization by modifying the above equation following the definition of spin polarization, $P_y = (I_\uparrow - I_\downarrow)/(I_\uparrow + I_\downarrow)$, as follows,
\begin{equation}
P_y = \frac{\cos^2{\theta} \cos^2{\alpha_0} - \sin^2{\alpha_0} + \gamma^2 \sin^2{\theta}\cos^2{\alpha_0} }{\cos^2{\theta} \cos^2{\alpha_0} + \sin^2{\alpha_0} + \gamma^2 \sin^2{\theta}\cos^2{\alpha_0}},
\end{equation}
where $A_z$ is given by $\sin{\theta}\cos{\alpha_0}$ in this geometry, and $\gamma^2$ is a fit parameter and measure of the relative contribution of $A_z$ to the total photoemission intensity ($\gamma$ corresponds to $2\beta/\alpha$ if we use the variables in Ref.~\onlinecite{CheolHwan}).
The final expression used to fit the data includes an overall coefficient (to account for finite resolution and other effects which can slightly reduce the measured polarizations in such spin-ARPES experiments) and a constant offset to $\alpha_0$ to allow for a slight misalignment of the photon polarization orientation.
The fit shown in Fig.~1(f) is obtained with a value of about 0.72 for the overall coefficient, an offset of $\alpha_0$ of $3.2^{\circ}$, and a $\gamma^2$ parameter of 2.0.

\section{$P_z$ maps of Bi$_2$Se$_3$ taken with circularly polarized light}

Again, following the approaches presented in the main text and Ref.~\onlinecite{CheolHwan}, circularly polarized light, at normal incidence to the surface of a topological insulator, will lead to a photoelectron spin texture oriented completely out-of-plane for a fixed $|P_z|$ at all $\mathbf{k}$, and with the sign of $P_z$ determined by the handedness of the light.
The experimentally measured $P_z$ maps shown in Figs.~3(c,d) of the main paper mostly agree with these predictions.
However, streaks of reduced $P_z$ follow the surface state dispersion along the right hand side at positive $k_x$.
This is likely due to the off-normal incidence of the light in the present experiment (see Fig.~S1), which will modify the expected spin polarization including reduced values of $P_z$.

Note that in the present geometry, negative values of $k_x$ (on the left side of Figs.~3(c,d)) are measured with the sample turned \textit{towards} the photons, and thus closer to normal incidence, while positive values of $k_x$ (on the right side of Figs.~3(c,d)) are measured with the sample turned \textit{away} from the photons, and thus further from normal incidence with lower values of $P_z$ expected.
Altering the experimental geometry, for instance to have the light at normal incidence, would provide direct insight into this issue.
Reduced values of $P_z$ may also be due to the light not being fully circularly polarized.  Imperfections in the photon polarization may be introduced by slight birefringence in the vacuum chamber's fused silica window, induced by mechanical and thermal stress associated with installation and bakeout.

\section{Spin-degeneracy of final states}

We note that the theory in Ref.~\onlinecite{CheolHwan} assumes spin-degenerate final states.
Even if this assumption is modified, it is reasonable to expect that the spin polarization of the photoelectrons could still be different from that of the originating surface states and dependent on the photon polarization, albeit in a manner somewhat different from that predicted by the theory in Ref.~\onlinecite{CheolHwan}.
The spin-degeneracy of the final states is not a strictly necessary condition to observe novel photoelectron spin-flipping such as observed in the current experiment.
On the other hand, because ARPES circular dichroism experiments on Bi$_2$Se$_3$ with similar photon energy are well described by theory based on the similar assumption of spin-degenerate final states,\cite{Wang2011b} the assumption is likely valid for the experimental results shown in the present work.

\section{Circular dichroism in Bi$_2$Se$_3$}

It was previously shown in Bi$_2$Se$_3$ that the total (spin-integrated) surface state photoemission intensity in ARPES data taken with circularly polarized light is sensitive to the relative alignment of photon helicity and surface state spin orientation.\cite{Wang2011b}
This was argued to provide indirect experimental access to the spin texture of the helical Dirac surface states.
Such studies, however, do not resolve the actual spin polarization of the photoexcited electrons.
Our current observations constitute an entirely different facet of the physics involved.

\section{Comparison of Bi$_2$Se$_3$ and Au(111)}

The spin-resolved ARPES from Bi$_2$Se$_3$ and Au(111) present an interesting juxtaposition.
In the present geometries, the spin polarization of photoelectrons from the Bi$_2$Se$_3$ topological surface state shows an extremely strong dependence on the photon polarization.
The top row of Fig.~S4 presents a couple of examples.
Panel (b) shows the measured $y$ component of photoelectron spin polarization, $P_y$, corresponding to the three vertical cuts marked in panel (a).
The $P_y$ curves are shown for both $p$- and $s$-polarized light.
The data corresponds to the full maps shown in Figs.~2(c)~and~(d) of the main paper, but offers a different view by directly comparing a few individual curves.
There is an enormous change in $P_y$ between the two photon polarizations.
In line with the discussion of Section SI II, there are two components to the photon polarization dependence of the $P_y$ curves visible here.
The $P_y$ curve corresponding to $\theta=0^{\circ}$ ($k_x=0$) shows moderately positive $P_y$ with $p$-polarized light, and nearly zero $P_y$ for $s$-polarized light.
This is due to the SME-induced effect discussed above.
As the SME effects are effectively $k$-independent, this contribution to the photon polarization dependence should be $k$ ($\theta$) independent.
The $P_y$ curves at $\theta=\pm 9^{\circ}$ ($k_x=\pm k_\textrm{F}$), however, show opposing behaviors.
At $\theta=-9^{\circ}$, $P_y$ near $E_\textrm{F}$ measured with $P$-polarized light is more positive than that measured with $s$-polarized light, while the opposite is true at $\theta=+9^{\circ}$.
This $k$-dependent contribution to the photon polarization dependence reflects the behavior of helical Dirac fermions as presented in the main manuscript.

\begin{figure*} \includegraphics[width=15cm]{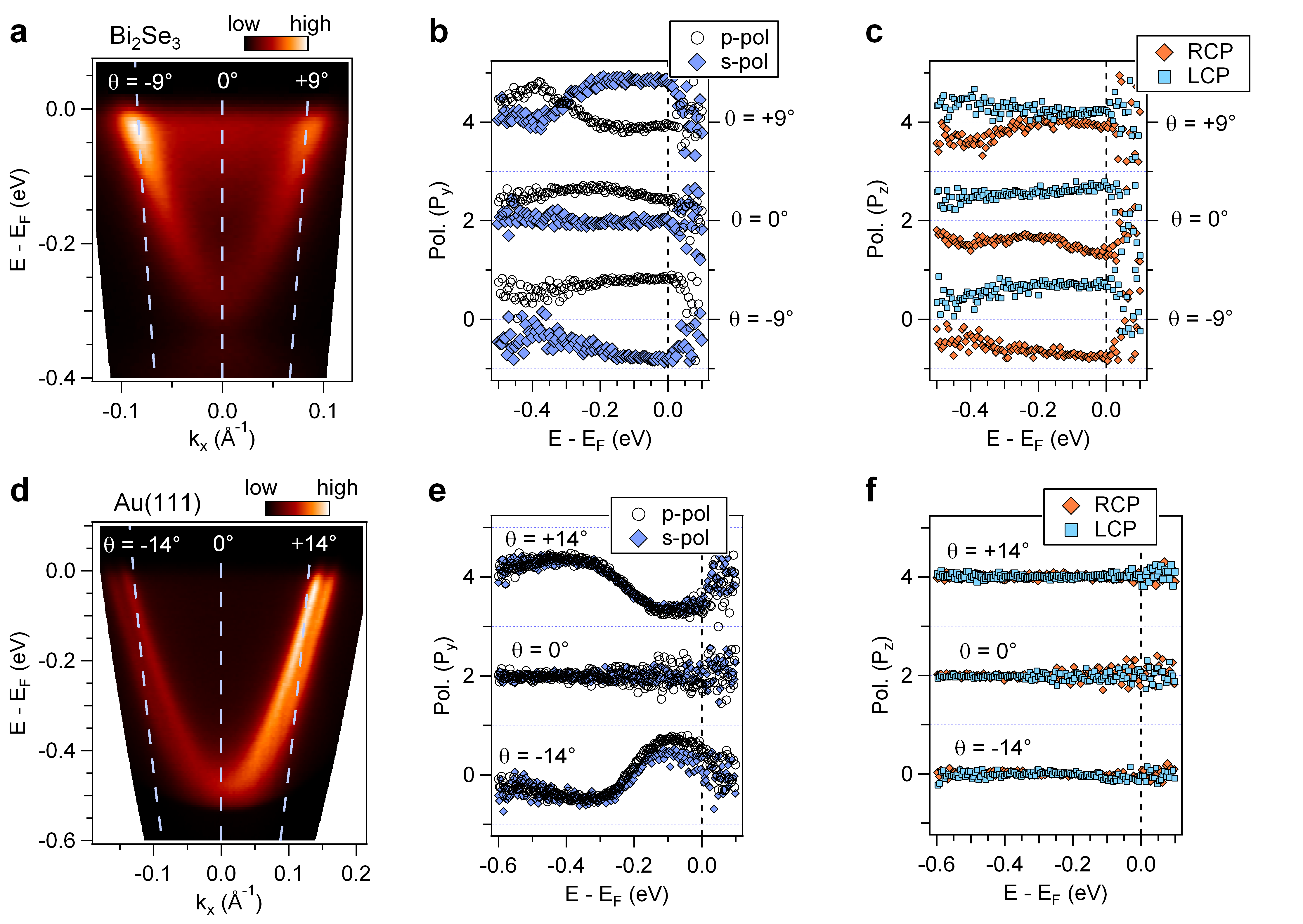}
\caption{\label{fig:fig1} \textbf{Polarization dependence of photoelectron spin in Bi$_2$Se$_3$ and Au(111) surface states.}  
(a) ARPES intensity map of Bi$_2$Se$_3$ as a function of binding energy and momentum, along $\Gamma$M, taken with $h\nu=6$ eV.
(b) The $y$ component of photoelectron spin polarization, $P_y$, as a function of binding energy at labeled emission angles, corresponding to the line cuts marked in (a).  The $P_y$ curves at each emission angle are vertically offset by 2 for clarity.  Corresponding $P_y$ curves measured with $p$- and $s$-polarized light are directly compared.
(c) Same as (b), except the curves display the $z$ component of photoelectron spin polarization, $P_z$, and are measured with both helicities of circular polarized light. 
(d-f) Same as (a-c) but measured from the Au(111) Shockley surface state.
}
\end{figure*}

Figure~S4(c) shows similar data, displaying the $z$ component, $P_z$, taken with right- and left-hand circularly polarized light.
This data corresponds to the full maps shown in Figs.~3(c)~and~(d) of the main paper.
Again, there is a very large dependence change with photon polarization, in line with Ref.~\onlinecite{CheolHwan}.

The bottom row of Fig.~S4 presents the corresponding data for the Au(111) surface state.
Although this surface state has a similar helical spin-texture, also as a result of spin-orbit coupling, the data shows a nearly constant photoelectron spin polarization at each photon polarization, in striking contrast with Bi$_2$Se$_3$.
This is true for $P_y$ (Figs.~S4(b)~and (e), and Fig.~2 of the main paper) and $P_z$ (Figs.~S4(c)~and~(f)).
Panel~(e) also shows that there is no $k$-independent component induced by SME effects ($P_y$ at $\theta=0$, or $k_x=0$, is zero for both $s$- and $p$-polarized light).

This contrasting behavior may be related to the contrasting circular dichroism in standard spin-integrated ARPES experiments on Bi$_2$Se$_3$ and Au.
In the case of Bi$_2$Se$_3$, a strong circular dichroism in the ARPES signal (difference in photoelectron intensity when illuminated with right- and left-hand circularly polarized light) was observed that has a texture in momentum-space which closely matches that of the surface state's spin texture.\cite{Wang2011b,Park2012a}
However, in the case of Au, a strong circular dichroism was observed that has a texture in momentum space that does not at all match that of the surface state's spin texture.\cite{Kim2012}

Kim~\textit{et al.}\cite{Kim2012} discusses these results in terms of the relative strengths of two particular terms in the Hamiltonian: `$H_\textrm{SOC}$', a term due to spin-orbit coupling, and `$H_\textrm{ES}$', a term due to the inversion symmetry breaking electrostatic field at the surface.
It is argued that in Bi$_2$Se$_3$, the $H_\textrm{SOC}$ term dominates, and leads to the circular dichroism behavior that mirrors the surface state spin texture.
In the Au, however, it is argued that relatively weaker spin-orbit coupling allows the $H_\textrm{ES}$ term to dominate, which leads to the contrasting circular dichroism that does not mirror the surface state spin texture.
This scenario may also apply to our current observations.
The $H_\textrm{SOC}$ term is the spin-flip term in the photoemission interaction Hamiltonian, as described in the main text, and its dominance in Bi$_2$Se$_3$ is consistent with the observed photoelectron spin-flipping.
The $H_\textrm{ES}$ term, in contrast, leads to a spin-conserving term in the photoemission interaction Hamiltonian.
Thus the weaker spin-orbit coupling and relative dominance of $H_\textrm{ES}$ in Au could explain the observed lack of photoelectron spin-flipping.

Similar experiments on other systems with varying and intermediate spin-orbit coupling strengths, such as BiTl(S$_1-\delta$Se$_{\delta}$)$_2$\cite{Xu2011a} and the adsorbate-induced Rashba states on Bi$_2$Se$_3$,\cite{King2011,Bahramy2012a} would be helpful in investigating this picture.
We note that following Ref.~\onlinecite{Kim2012}, the $H_\textrm{ES}$ term is linearly dependent on $k$.
In this scenario, apparently this term remains dominant to the spin-orbit term down to momentum values below which could be experimentally resolved (in either our spin-resolved measurement, or the spin-integrated circular dichroism experiment\cite{Kim2012}).
This overall picture may be supported if a reversal of behavior, from Bi$_2$Se$_3$-like to Au-like, was observed as a function of increasing $k$ in a system with intermediate spin-orbit coupling.

\end{spacing}


%% file: Jozwiak_TI_total.bbl
\begin{thebibliography}{10}
\expandafter\ifx\csname url\endcsname\relax
  \def\url#1{\texttt{#1}}\fi
\expandafter\ifx\csname urlprefix\endcsname\relax\def\urlprefix{URL }\fi
\providecommand{\bibinfo}[2]{#2}
\providecommand{\eprint}[2][]{\url{#2}}

\bibitem{Fu2007}
\bibinfo{author}{Fu, L.}, \bibinfo{author}{Kane, C.~L.} \&
  \bibinfo{author}{Mele, E.~J.}
\newblock \bibinfo{title}{Topological insulators in three dimensions}.
\newblock \emph{\bibinfo{journal}{Phys. Rev. Lett.}}
  \textbf{\bibinfo{volume}{98}}, \bibinfo{pages}{106803}
  (\bibinfo{year}{2007}).

\bibitem{Moore2007}
\bibinfo{author}{Moore, J.~E.} \& \bibinfo{author}{Balents, L.}
\newblock \bibinfo{title}{Topological invariants of time-reversal-invariant
  band structures}.
\newblock \emph{\bibinfo{journal}{Phys. Rev. B}} \textbf{\bibinfo{volume}{75}},
  \bibinfo{pages}{121306} (\bibinfo{year}{2007}).

\bibitem{Roy2009}
\bibinfo{author}{Roy, R.}
\newblock \bibinfo{title}{Topological phases and the quantum spin Hall effect
  in three dimensions}.
\newblock \emph{\bibinfo{journal}{Phys. Rev. B}} \textbf{\bibinfo{volume}{79}},
  \bibinfo{pages}{195322--} (\bibinfo{year}{2009}).

\bibitem{Qi2010}
\bibinfo{author}{Qi, X.-L.} \& \bibinfo{author}{Zhang, S.-C.}
\newblock \bibinfo{title}{The quantum spin Hall effect and topological
  insulators}.
\newblock \emph{\bibinfo{journal}{Physics Today}}
  \textbf{\bibinfo{volume}{63}}, \bibinfo{pages}{33--38}
  (\bibinfo{year}{2010}).

\bibitem{Moore2010}
\bibinfo{author}{Moore, J.~E.}
\newblock \bibinfo{title}{The birth of topological insulators}.
\newblock \emph{\bibinfo{journal}{Nature}} \textbf{\bibinfo{volume}{464}},
  \bibinfo{pages}{194--198} (\bibinfo{year}{2010}).

\bibitem{Hsieh2009a}
\bibinfo{author}{Hsieh, D.} \emph{et~al.}
\newblock \bibinfo{title}{A tunable topological insulator in the spin helical
  Dirac transport regime}.
\newblock \emph{\bibinfo{journal}{Nature}} \textbf{\bibinfo{volume}{460}},
  \bibinfo{pages}{1101--1105} (\bibinfo{year}{2009}).

\bibitem{Hsieh2009}
\bibinfo{author}{Hsieh, D.} \emph{et~al.}
\newblock \bibinfo{title}{Observation of unconventional quantum spin textures
  in topological insulators}.
\newblock \emph{\bibinfo{journal}{Science}} \textbf{\bibinfo{volume}{323}},
  \bibinfo{pages}{919--922} (\bibinfo{year}{2009}).

\bibitem{Nishide2010}
\bibinfo{author}{Nishide, A.} \emph{et~al.}
\newblock \bibinfo{title}{Direct mapping of the spin-filtered surface bands of
  a three-dimensional quantum spin Hall insulator}.
\newblock \emph{\bibinfo{journal}{Phys. Rev. B}} \textbf{\bibinfo{volume}{81}},
  \bibinfo{pages}{041309} (\bibinfo{year}{2010}).

\bibitem{Pan2011a}
\bibinfo{author}{Pan, Z.-H.} \emph{et~al.}
\newblock \bibinfo{title}{Electronic structure of the topological insulator
  Bi$_2$Se$_3$ using angle-resolved photoemission spectroscopy: Evidence for a nearly
  full surface spin polarization}.
\newblock \emph{\bibinfo{journal}{Phys. Rev. Lett.}}
  \textbf{\bibinfo{volume}{106}}, \bibinfo{pages}{257004}
  (\bibinfo{year}{2011}).

\bibitem{Souma2011a}
\bibinfo{author}{Souma, S.} \emph{et~al.}
\newblock \bibinfo{title}{Direct measurement of the out-of-plane spin texture
  in the Dirac-cone surface state of a topological insulator}.
\newblock \emph{\bibinfo{journal}{Phys. Rev. Lett.}}
  \textbf{\bibinfo{volume}{106}}, \bibinfo{pages}{216803}
  (\bibinfo{year}{2011}).

\bibitem{Xu2011a}
\bibinfo{author}{Xu, S.-Y.} \emph{et~al.}
\newblock \bibinfo{title}{Topological phase transition and texture inversion in
  a tunable topological insulator}.
\newblock \emph{\bibinfo{journal}{Science}} \textbf{\bibinfo{volume}{332}},
  \bibinfo{pages}{560--564} (\bibinfo{year}{2011}).

\bibitem{Jozwiak2011}
\bibinfo{author}{Jozwiak, C.} \emph{et~al.}
\newblock \bibinfo{title}{Widespread spin polarization effects in photoemission
  from topological insulators}.
\newblock \emph{\bibinfo{journal}{Phys. Rev. B}} \textbf{\bibinfo{volume}{84}},
  \bibinfo{pages}{165113} (\bibinfo{year}{2011}).

\bibitem{Jozwiak2010}
\bibinfo{author}{Jozwiak, C.} \emph{et~al.}
\newblock \bibinfo{title}{A high-efficiency spin-resolved photoemission
  spectrometer combining time-of-flight spectroscopy with exchange-scattering
  polarimetry}.
\newblock \emph{\bibinfo{journal}{Rev. Sci. Instrum.}}
  \textbf{\bibinfo{volume}{81}}, \bibinfo{pages}{053904}
  (\bibinfo{year}{2010}).

\bibitem{Qi2008}
\bibinfo{author}{Qi, X.-L.}, \bibinfo{author}{Hughes, T.~L.} \&
  \bibinfo{author}{Zhang, S.-C.}
\newblock \bibinfo{title}{Topological field theory of time-reversal invariant
  insulators}.
\newblock \emph{\bibinfo{journal}{Phys. Rev. B}} \textbf{\bibinfo{volume}{78}},
  \bibinfo{pages}{195424} (\bibinfo{year}{2008}).

\bibitem{Fu2008}
\bibinfo{author}{Fu, L.} \& \bibinfo{author}{Kane, C.~L.}
\newblock \bibinfo{title}{Superconducting proximity effect and Majorana
  fermions at the surface of a topological insulator}.
\newblock \emph{\bibinfo{journal}{Phys. Rev. Lett.}}
  \textbf{\bibinfo{volume}{100}}, \bibinfo{pages}{096407}
  (\bibinfo{year}{2008}).

\bibitem{Nayak2008}
\bibinfo{author}{Nayak, C.}, \bibinfo{author}{Simon, S.~H.},
  \bibinfo{author}{Stern, A.}, \bibinfo{author}{Freedman, M.} \&
  \bibinfo{author}{Das~Sarma, S.}
\newblock \bibinfo{title}{Non-Abelian anyons and topological quantum
  computation}.
\newblock \emph{\bibinfo{journal}{Rev. Mod. Phys.}}
  \textbf{\bibinfo{volume}{80}}, \bibinfo{pages}{1083} (\bibinfo{year}{2008}).

\bibitem{Roushan2009}
\bibinfo{author}{Roushan, P.} \emph{et~al.}
\newblock \bibinfo{title}{Topological surface states protected from
  backscattering by chiral spin texture}.
\newblock \emph{\bibinfo{journal}{Nature}} \textbf{\bibinfo{volume}{460}},
  \bibinfo{pages}{1106--1109} (\bibinfo{year}{2009}).

\bibitem{Peng2010}
\bibinfo{author}{Peng, H.} \emph{et~al.}
\newblock \bibinfo{title}{Aharonov-Bohm interference in topological insulator
  nanoribbons}.
\newblock \emph{\bibinfo{journal}{Nat Mater}} \textbf{\bibinfo{volume}{9}},
  \bibinfo{pages}{225--229} (\bibinfo{year}{2010}).

\bibitem{Hsieh2011}
\bibinfo{author}{Hsieh, D.} \emph{et~al.}
\newblock \bibinfo{title}{Nonlinear optical probe of tunable surface electrons
  on a topological insulator}.
\newblock \emph{\bibinfo{journal}{Phys. Rev. Lett.}}
  \textbf{\bibinfo{volume}{106}}, \bibinfo{pages}{057401}
  (\bibinfo{year}{2011}).

\bibitem{McIver2012}
\bibinfo{author}{McIver, J.~W.}, \bibinfo{author}{Hsieh, D.},
  \bibinfo{author}{Steinberg, H.}, \bibinfo{author}{Jarillo-Herrero, P.} \&
  \bibinfo{author}{Gedik, N.}
\newblock \bibinfo{title}{Control over topological insulator photocurrents with
  light polarization}.
\newblock \emph{\bibinfo{journal}{Nat Nano}} \textbf{\bibinfo{volume}{7}},
  \bibinfo{pages}{96--100} (\bibinfo{year}{2012}).

\bibitem{Zhang2009}
\bibinfo{author}{Zhang, H.} \emph{et~al.}
\newblock \bibinfo{title}{Topological insulators in Bi$_2$Se$_3$, Bi$_2$Te$_3$ and Sb$_2$Te$_3$
  with a single Dirac cone on the surface}.
\newblock \emph{\bibinfo{journal}{Nat Phys}} \textbf{\bibinfo{volume}{5}},
  \bibinfo{pages}{438--442} (\bibinfo{year}{2009}).

\bibitem{CheolHwan}
\bibinfo{author}{Park, C.-H.} \& \bibinfo{author}{Louie, S.~G.}
\newblock \bibinfo{title}{Spin polarization of photoelectrons from topological
  insulators}.
\newblock \emph{\bibinfo{journal}{Phys. Rev. Lett.}}
  \textbf{\bibinfo{volume}{109}}, \bibinfo{pages}{097601}
  (\bibinfo{year}{2012}).

\bibitem{Xia2009}
\bibinfo{author}{Xia, Y.} \emph{et~al.}
\newblock \bibinfo{title}{Observation of a large-gap topological-insulator
  class with a single Dirac cone on the surface}.
\newblock \emph{\bibinfo{journal}{Nat Phys}} \textbf{\bibinfo{volume}{5}},
  \bibinfo{pages}{398--402} (\bibinfo{year}{2009}).

\bibitem{Hoesch2004}
\bibinfo{author}{Hoesch, M.} \emph{et~al.}
\newblock \bibinfo{title}{Spin structure of the Shockley surface state on
  Au(111)}.
\newblock \emph{\bibinfo{journal}{Phys. Rev. B}} \textbf{\bibinfo{volume}{69}},
  \bibinfo{pages}{241401} (\bibinfo{year}{2004}).

\bibitem{Mirhosseini2012}
\bibinfo{author}{Mirhosseini, H.} \& \bibinfo{author}{Henk, J.}
\newblock \bibinfo{title}{Spin texture and circular dichroism in photoelectron
  spectroscopy from the topological insulator Bi$_{2}$Te$_{3}$:
  First-principles photoemission calculations}.
\newblock \emph{\bibinfo{journal}{Phys. Rev. Lett.}}
  \textbf{\bibinfo{volume}{109}}, \bibinfo{pages}{036803}
  (\bibinfo{year}{2012}).

\bibitem{Henk2003}
\bibinfo{author}{Henk, J.}, \bibinfo{author}{Ernst, A.} \&
  \bibinfo{author}{Bruno, P.}
\newblock \bibinfo{title}{Spin polarization of the L-gap surface states on
  Au(111)}.
\newblock \emph{\bibinfo{journal}{Phys. Rev. B}} \textbf{\bibinfo{volume}{68}},
  \bibinfo{pages}{165416} (\bibinfo{year}{2003}).

\bibitem{Kim2012}
\bibinfo{author}{Kim, B.} \emph{et~al.}
\newblock \bibinfo{title}{Spin and orbital angular momentum structure of
  Cu(111) and Au(111) surface states}.
\newblock \emph{\bibinfo{journal}{Phys. Rev. B}} \textbf{\bibinfo{volume}{85}},
  \bibinfo{pages}{195402} (\bibinfo{year}{2012}).

\bibitem{Pierce1980}
\bibinfo{author}{Pierce, D.~T.} \emph{et~al.}
\newblock \bibinfo{title}{The GaAs spin polarized electron source}.
\newblock \emph{\bibinfo{journal}{Rev. Sci. Instrum.}}
  \textbf{\bibinfo{volume}{51}}, \bibinfo{pages}{478} (\bibinfo{year}{1980}).

\bibitem{King2011}
\bibinfo{author}{King, P. D.~C.} \emph{et~al.}
\newblock \bibinfo{title}{Large tunable Rashba spin splitting of a
  two-dimensional electron gas in Bi$_{2}$Se$_{3}$}.
\newblock \emph{\bibinfo{journal}{Phys. Rev. Lett.}}
  \textbf{\bibinfo{volume}{107}}, \bibinfo{pages}{096802}
  (\bibinfo{year}{2011}).

\bibitem{Ishizaka2011}
\bibinfo{author}{Ishizaka, K.} \emph{et~al.}
\newblock \bibinfo{title}{Giant Rashba-type spin splitting in bulk BiTeI}.
\newblock \emph{\bibinfo{journal}{Nat Mater}} \textbf{\bibinfo{volume}{10}},
  \bibinfo{pages}{521--526} (\bibinfo{year}{2011}).

\end{thebibliography}

\begin{thebibliography}{32}
\expandafter\ifx\csname natexlab\endcsname\relax\def\natexlab#1{#1}\fi
\expandafter\ifx\csname url\endcsname\relax
  \def\url#1{\texttt{#1}}\fi
\expandafter\ifx\csname urlprefix\endcsname\relax\def\urlprefix{URL }\fi

\large


\bibitem[{Jozwiak \emph{et~al.}(2010)}]{Jozwiak2010}
Jozwiak, C. \emph{et~al.}
\newblock A high-efficiency spin-resolved photoemission spectrometer combining
  time-of-flight spectroscopy with exchange-scattering polarimetry.
\newblock \emph{Rev. Sci. Instrum.} \textbf{81}, 053904 (2010).

\bibitem[{Fano(1969)}]{Fano1969}
Fano, U.
\newblock Spin orientation of photoelectrons ejected by circularly polarized
  light.
\newblock \emph{Phys. Rev.} \textbf{178}, 131 (1969).

\bibitem[{Lee(1974)}]{Lee1974}
Lee, C.~M.
\newblock Spin polarization and angular distribution of photoelectrons in the
  Jacob-Wick helicity formalism. Application to autoionization resonances.
\newblock \emph{Phys. Rev. A} \textbf{10}, 1598 (1974).

\bibitem[{Cherepkov(1979)}]{Cherepkov1979}
Cherepkov, N.~A.
\newblock Spin polarisation of photoelectrons ejected from unpolarised atoms.
\newblock \emph{Journal of Physics B: Atomic and Molecular Physics}
  \textbf{12}, 1279 (1979).

\bibitem[{Kessler(1976)}]{Kesslerbook}
Kessler, J.
\newblock \emph{Polarized Electrons} (Springer, 1976).

\bibitem[{Heinzmann \emph{et~al.}(1979)Heinzmann, Sch\"onhense \&
  Kessler}]{Heinzmann1979}
Heinzmann, U., Sch\"onhense, G. \& Kessler, J.
\newblock Polarization of photoelectrons ejected by unpolarized light from
  xenon atoms.
\newblock \emph{Phys. Rev. Lett.} \textbf{42}, 1603 (1979).

\bibitem[{Roth \emph{et~al.}(1994)Roth, Hillebrecht, Park, Rose \&
  Kisker}]{Roth1994}
Roth, C., Hillebrecht, F.~U., Park, W.~G., Rose, H.~B. \& Kisker, E.
\newblock Spin polarization in Cu core-level photoemission with linearly
  polarized soft X-rays.
\newblock \emph{Phys. Rev. Lett.} \textbf{73}, 1963 (1994).

\bibitem[{Rose \emph{et~al.}(1996)Rose, Fanelsa, Kinoshita, Roth, Hillebrecht
  \& Kisker}]{Rose1996}
Rose, H.~B., Fanelsa, A., Kinoshita, T., Roth, C., Hillebrecht, F.~U. \&
  Kisker, E.
\newblock Spin-orbit-induced spin polarization in W 4f photoemission.
\newblock \emph{Phys. Rev. B} \textbf{53}, 1630 (1996).

\bibitem[{Yu \& Tobin(2008)}]{Yu2008}
Yu, S.-W. \& Tobin, J.~G.
\newblock Breakdown of spatial inversion symmetry in core-level photoemission
  of Pt(001).
\newblock \emph{Phys. Rev. B} \textbf{77}, 193409 (2008).

\bibitem[{Jozwiak \emph{et~al.}(2011)}]{Jozwiak2011}
Jozwiak, C. \emph{et~al.}
\newblock Widespread spin polarization effects in photoemission from
  topological insulators.
\newblock \emph{Phys. Rev. B} \textbf{84}, 165113 (2011).

\bibitem[{Tamura \emph{et~al.}(1987)Tamura, Piepke \& Feder}]{Tamura1987}
Tamura, E., Piepke, W. \& Feder, R.
\newblock New spin-polarization effect in photoemission from nonmagnetic
  surfaces.
\newblock \emph{Phys. Rev. Lett.} \textbf{59}, 934 (1987).

\bibitem[{Tamura \& Feder(1991{\natexlab{a}})}]{Tamura1991}
Tamura, E. \& Feder, R.
\newblock Spin-polarized normal photoemission from non-magnetic (111)-surfaces
  by p-polarized light.
\newblock \emph{Solid State Communications} \textbf{79}, 989--993
  (1991{\natexlab{a}}).

\bibitem[{Tamura \& Feder(1991{\natexlab{b}})}]{Tamura1991a}
Tamura, E. \& Feder, R.
\newblock Spin polarization in normal photoemission by linearly polarized light
  from nonmagnetic (001) surfaces.
\newblock \emph{Europhys. Lett.} \textbf{16}, 695 (1991{\natexlab{b}}).

\bibitem[{Henk \& Feder(1994)}]{Henk1994}
Henk, J. \& Feder, R.
\newblock Spin polarization in normal photoemission by linearly polarized light
  from non-magnetic (110) surfaces.
\newblock \emph{Europhys. Lett.} \textbf{28}, 609 (1994).

\bibitem[{Schmiedeskamp \emph{et~al.}(1988)Schmiedeskamp, Vogt \&
  Heinzmann}]{Schmiedeskamp1988}
Schmiedeskamp, B., Vogt, B. \& Heinzmann, U.
\newblock Experimental verification of a new spin-polarization effect in
  photoemission: Polarized photoelectrons from Pt(111) with linearly polarized
  radiation in normal incidence and normal emission.
\newblock \emph{Phys. Rev. Lett.} \textbf{60}, 651 (1988).

\bibitem[{Schmiedeskamp \emph{et~al.}(1991)Schmiedeskamp, Irmer, David \&
  Heinzmann}]{Schmiedeskamp1991}
Schmiedeskamp, B., Irmer, N., David, R. \& Heinzmann, U.
\newblock A new spin effect in photoemission with unpolarized light:
  Experimental evidence of spin polarized electrons in normal emission from
  Pt(111) and Au(111).
\newblock \emph{Applied Physics A} \textbf{53}, 418 (1991).

\bibitem[{Irmer \emph{et~al.}(1992)Irmer, David, Schmiedeskamp \&
  Heinzmann}]{Irmer1992}
Irmer, N., David, R., Schmiedeskamp, B. \& Heinzmann, U.
\newblock Experimental verification of a spin effect in photoemission:
  Polarized electrons due to phase-shift differences in the normal emission
  from Pt(100) by unpolarized radiation.
\newblock \emph{Phys. Rev. B} \textbf{45}, 3849 (1992).

\bibitem[{Irmer \emph{et~al.}(1994)Irmer, Frentzen, Schmiedeskamp \&
  Heinzmann}]{Irmer1994}
Irmer, N., Frentzen, F., Schmiedeskamp, B. \& Heinzmann, U.
\newblock Spin polarized photoelectrons with unpolarized light in normal
  emission from Pt(110).
\newblock \emph{Surface Science} \textbf{307-309}, 1114--1117 (1994).

\bibitem[{Irmer \emph{et~al.}(1995)Irmer, Frentzen, David, Stoppmanns,
  Schmiedeskamp \& Heinzmann}]{Irmer1995}
Irmer, N., Frentzen, F., David, R., Stoppmanns, P., Schmiedeskamp, B. \&
  Heinzmann, U.
\newblock Photon energy dependence of spin-resolved photoemission spectra in
  normal emission from Pt(110) by linearly polarized light.
\newblock \emph{Surface Science} \textbf{331-333}, 1147--1151 (1995).

\bibitem[{Irmer \emph{et~al.}(1996)Irmer, Frentzen, Yu, Schmiedeskamp \&
  Heinzmann}]{Irmer1996}
Irmer, N., Frentzen, F., Yu, S.~W., Schmiedeskamp, B. \& Heinzmann, U.
\newblock A new effect in spin-resolved photoemission from Pt(110) in normal
  emission by linearly polarized VUV-radiation.
\newblock \emph{Journal of Electron Spectroscopy and Related Phenomena}
  \textbf{78}, 321--324 (1996).

\bibitem[{Kessler \& Lorenz(1970)}]{Kessler1970}
Kessler, J. \& Lorenz, J.
\newblock Experimental Verification of the Fano effect.
\newblock \emph{Phys. Rev. Lett.} \textbf{24}, 87 (1970).

\bibitem[{Starke \emph{et~al.}(1996)}]{Starke1996}
Starke, K. \emph{et~al.}
\newblock Spin-polarized photoelectrons excited by circularly polarized
  radiation from a nonmagnetic solid.
\newblock \emph{Phys. Rev. B} \textbf{53}, R10544--R10547 (1996).

\bibitem[{Feuchtwang \emph{et~al.}(1978)Feuchtwang, Cutler \&
  Nagy}]{Feuchtwang1978a}
Feuchtwang, T., Cutler, P. \& Nagy, D.
\newblock A review of the theoretical and experimental analyses of electron
  spin polarization in ferromagnetic transition metals: II. New theoretical
  results for the analysis of ESP in field emission, photoemission, and
  tunneling.
\newblock \emph{Surface Science} \textbf{75}, 490--528 (1978).

\bibitem[{Feder(1985)}]{Federbook}
Feder, R. (ed.).
\newblock \emph{Polarized electrons in surface physics} (World Scientific,
  1985).

\bibitem[{Park \& Louie(2012)}]{CheolHwan}
Park, C.-H. \& Louie, S.~G.
\newblock Spin polarization of photoelectrons from topological insulators.
\newblock \emph{Phys. Rev. Lett.} \textbf{109}, 097601 (2012).

\bibitem[{Fu(2009)}]{Fu2009}
Fu, L.
\newblock Hexagonal warping effects in the surface states of the topological
  insulator Bi$_{2}$Te$_{3}$.
\newblock \emph{Phys. Rev. Lett.} \textbf{103}, 266801 (2009).

\bibitem[{Wang \emph{et~al.}(2011)Wang, Hsieh, Pilon, Fu, Gardner, Lee \&
  Gedik}]{Wang2011b}
Wang, Y.~H., Hsieh, D., Pilon, D., Fu, L., Gardner, D.~R., Lee, Y.~S. \& Gedik,
  N.
\newblock Observation of a warped helical spin texture in Bi$_{2}$Se$_{3}$ from
  circular dichroism angle-resolved photoemission spectroscopy.
\newblock \emph{Phys. Rev. Lett.} \textbf{107}, 207602 (2011).

\bibitem[{Park \emph{et~al.}(2012)}]{Park2012a}
Park, S.~R. \emph{et~al.}
\newblock Chiral orbital-angular momentum in the surface states of
  Bi$_{2}$Se$_{3}$.
\newblock \emph{Phys. Rev. Lett.} \textbf{108}, 046805 (2012).

\bibitem[{Kim \emph{et~al.}(2012)}]{Kim2012}
Kim, B. \emph{et~al.}
\newblock Spin and orbital angular momentum structure of Cu(111) and Au(111)
  surface states.
\newblock \emph{Phys. Rev. B} \textbf{85}, 195402 (2012).

\bibitem[{Xu \emph{et~al.}(2011)}]{Xu2011a}
Xu, S.-Y. \emph{et~al.}
\newblock Topological phase transition and texture tnversion in a tunable
  topological insulator.
\newblock \emph{Science} \textbf{332}, 560--564 (2011).

\bibitem[{King \emph{et~al.}(2011)}]{King2011}
King, P. D.~C. \emph{et~al.}
\newblock Large tunable Rashba spin splitting of a two-dimensional electron gas
  in Bi$_{2}$Se$_{3}$.
\newblock \emph{Phys. Rev. Lett.} \textbf{107}, 096802 (2011).

\bibitem[{Bahramy \emph{et~al.}(2012)}]{Bahramy2012a}
Bahramy, M. \emph{et~al.}
\newblock Emergent quantum confinement at topological insulator surfaces.
\newblock \emph{Nat Commun} \textbf{3}, 1159 (2012).


\end{thebibliography}
